\documentclass{article}

\usepackage{PRIMEarxiv}

\usepackage[utf8]{inputenc} 
\usepackage[T1]{fontenc}    
\usepackage{hyperref}       
\usepackage{url}            
\usepackage{booktabs}       
\usepackage{amsfonts}       
\usepackage{microtype}      
\usepackage{fancyhdr}       
\usepackage{graphicx}       
\graphicspath{{./}{media/}}
\usepackage{xcolor}

\title{Beyond Missing Data: Questionnaire Response Uncertainty as an Early Digital Sign of Dementia-Related Vulnerability
}

\author{
  \parbox[t]{0.44\textwidth}{\centering
    Yukun Lu\textsuperscript{$\dagger$}\\[0.4em]
    {\normalfont\small Shanghai Institute for Mathematics and Interdisciplinary Sciences, Shanghai 200433, China}\\[0.3em]
    {\normalfont\small Department of Statistics and Data Science, National University of Singapore, Singapore 117546}\\[0.3em]
    {\normalfont\small\texttt{lykun@simis.cn}}}
  \And
  \parbox[t]{0.44\textwidth}{\centering
    Bingjie Li\textsuperscript{$\dagger$}\\[0.4em]
    {\normalfont\small Department of Epidemiology and Biostatistics, School of Public Health, Tongji Medical College, Huazhong University of Science and Technology, Wuhan 430030, China}\\[0.3em]
    {\normalfont\small Department of Statistics and Data Science, National University of Singapore, Singapore 117546}\\[0.3em]
    {\normalfont\small\texttt{bingjieli@hust.edu.cn}}}
  \AND
  \parbox[t]{0.44\textwidth}{\centering
    Yuqing Xia\textsuperscript{*}\\[0.4em]
    {\normalfont\small School of Data Sciences, Zhejiang University of Finance and Economics, Hangzhou 310018, China}\\[0.3em]
    {\normalfont\small\texttt{yuqing.xia@zufe.edu.cn}}}
  \And
  \parbox[t]{0.44\textwidth}{\centering
    Zhigang Yao\textsuperscript{*}\\[0.4em]
    {\normalfont\small Department of Statistics and Data Science, National University of Singapore, Singapore 117546}\\[0.3em]
    {\normalfont\small\texttt{zhigang.yao@nus.edu.sg}}}
  \AND
  \parbox[t]{0.9\textwidth}{\centering\normalfont\small
    \textsuperscript{$\dagger$}Yukun Lu and Bingjie Li contributed equally.\\[0.2em]
    \textsuperscript{*}Correspondence to Yuqing Xia and Zhigang Yao.}
}

\begin{document}
\maketitle

\begin{abstract}
Explicit ``Do not know'' responses are treated as missing data, although their accumulation may encode uncertainty. We analysed 502,234 UK Biobank participants to test associations between response uncertainty and incident dementia-related diagnoses. Compared with participants giving 0--1 such responses, those giving more than seven had higher risks of the dementia-related disease composite (hazard ratio 1.62, 95\% confidence interval 1.50--1.74), Alzheimer's disease (1.63, 1.48--1.79), and vascular dementia (1.63, 1.42--1.86) after multivariable adjustment. Associations were approximately linear, persisted after a two-year lag, and were directionally reproduced in Scotland and Wales. Polygenic risk scores minimally attenuated these associations. Domain, molecular, and chronic-disease analyses placed response uncertainty within a broader multisystem vulnerability phenotype. Here, we show that explicit response uncertainty provides a scalable behavioural marker of subsequent dementia-related risk.
\end{abstract}


\section*{Introduction}

Dementia is one of the defining public-health challenges of an ageing population. An estimated 57.4 million people were living with dementia worldwide in 2019, and this number is projected to reach 152.8 million by 2050 \cite{gbd2019dementia}. Alzheimer's disease (AD), other neurodegenerative processes, and cerebrovascular pathology account for much of this burden, with consequences for affected individuals, families, health-care systems, and society \cite{livingston2020dementia,van2005epidemiology}. The pathological processes underlying dementia can precede clinical diagnosis by many years. Identifying vulnerability before overt cognitive and functional impairment is therefore important for longitudinal monitoring and prevention research. Cerebrospinal-fluid measures, positron-emission tomography, and magnetic-resonance imaging provide valuable biological and structural information \cite{jack2010hypothetical,sperling2011toward,dubois2016}. However, these approaches can be invasive, costly, resource-intensive, or difficult to deploy at population scale. Complementary signals already generated during routine health assessments could extend the reach of established approaches at low marginal cost.

Large-scale health studies routinely collect extensive questionnaire data, but the response process itself is rarely treated as a source of health information. In the UK Biobank touchscreen questionnaire, the explicit ``Do not know'' option (hereafter, a DK response) is encoded as a negative value and is usually treated as a missing observation \cite{ukb_touchscreen}. It may subsequently be handled through complete-case analysis, exclusion, missing-indicator approaches, or statistical imputation \cite{sterne2009multiple,pedersen2017missing,little2019statistical}. These methods are essential when the inferential target is the unanswered questionnaire item. Nevertheless, an explicit uncertain response remains an observed behaviour generated during the assessment. Producing a response requires comprehension of the question, retrieval of relevant information, evaluation of whether that information is sufficient, and selection of an answer. A DK response can therefore arise from insufficient knowledge, ambiguous wording, reluctance to disclose, reduced engagement, difficulty retrieving information, or uncertainty when integrating available information \cite{beatty1998dontknow}. A single DK response cannot be interpreted as cognitive impairment. Its accumulation across heterogeneous questions may nonetheless capture a reproducible pattern of response uncertainty. Earlier survey research in older adults linked item non-response and inconsistent responding with cognitive, physical, and psychological functioning \cite{colsher1989data}.

Longitudinal evidence strengthens the possibility that questionnaire behaviour may precede clinically recognized cognitive decline. In the Health and Retirement Study, subtle reporting irregularities, including incompatible, implausible, skipped, or internally inconsistent responses, were prospectively associated with dementia incidence \cite{schneider2021subtle}. Related analyses connected lower-quality questionnaire responding with poorer concurrent cognition and faster cognitive decline \cite{schneider2022quality}. An individual-participant-data meta-analysis across ten ageing cohorts subsequently found broadly consistent associations between cognitive functioning and several indicators of survey response quality \cite{schneider2024cognitive}. These observations complement research on subjective cognitive decline and altered cognitive self-appraisal \cite{jessen2014conceptual,jessen2020characterisation}, but response uncertainty represents a distinct source of information. It arises incidentally while an individual answers particular questions, rather than through an explicit report of worsening cognition. When captured electronically, cumulative DK responses form a scalable digital behavioural feature \cite{coravos2019digital}.

Important questions remain unresolved. Previous studies have generally examined composite indicators of erroneous or low-quality responding, whereas a cumulative DK count is simple, interpretable, and available in many cohorts. It is unknown whether DK burden shows a graded association with incident dementia-related outcomes after accounting for demographic, socioeconomic, behavioural, and clinical differences. It is also unclear whether the association persists after excluding diagnoses close to baseline or reproduces across geographic subsets. DK responses to household, medical, lifestyle, and personal-history questions may not carry equivalent information, making questionnaire-domain decomposition important. It also remains unknown whether questionnaire response uncertainty largely reflects measured polygenic susceptibility to dementia and vascular disease, or whether it provides complementary information beyond available polygenic risk scores (PRSs). Finally, circulating molecular profiles and non-neurological outcomes can establish whether DK burden occurs within a narrow dementia-related process or a broader phenotype of health vulnerability. Addressing these questions requires triangulation across temporal analyses, questionnaire domains, population subgroups, biological measurements, polygenic susceptibility, and geographically distinct samples.

\begin{figure*}[!t]
    \centering
    \includegraphics[width=\textwidth]{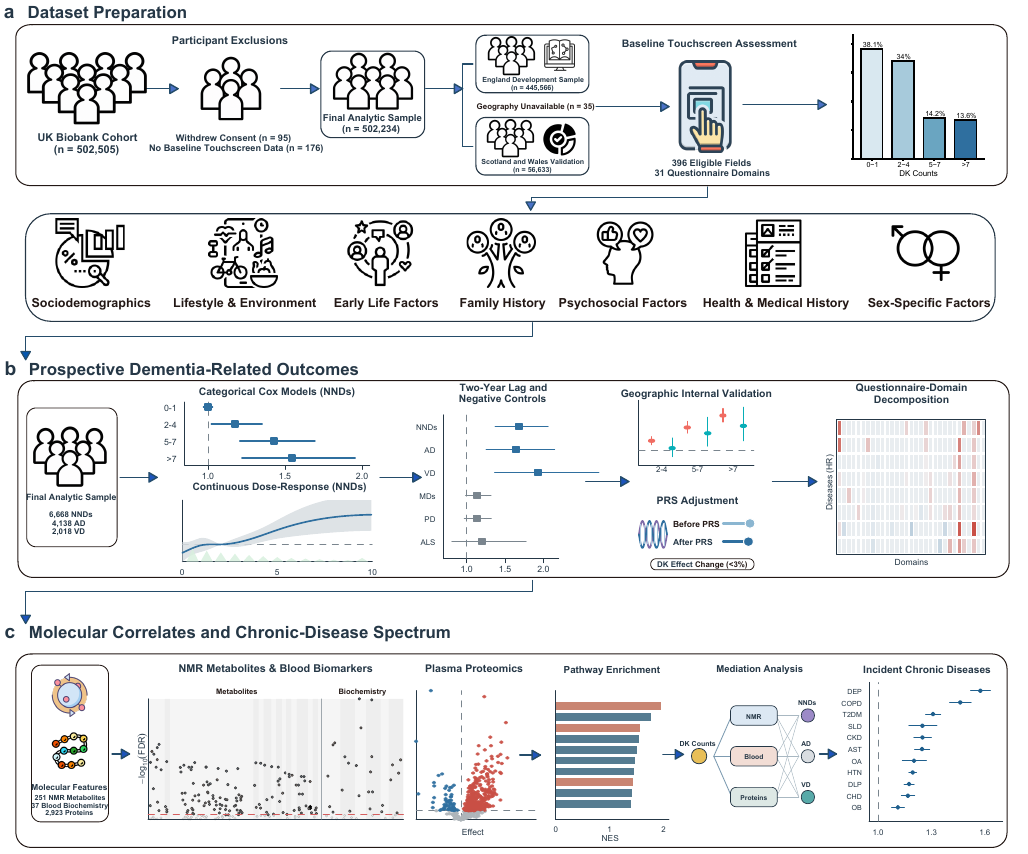}
    \caption{\textbf{Study design and analytical framework for evaluating DK
    response burden as an early digital sign of dementia-related
    vulnerability.} \textbf{a}, Construction of the analytic cohort,
    geographic division into the England development sample and the Scotland
    and Wales internal replication sample, assessment of 396 eligible
    touchscreen fields across 31 directly answered domains, and distribution of
    participants across the four DK response groups. \textbf{b}, Prospective
    analyses of dementia-related outcomes, including categorical and
    continuous Cox models, two-year lag and motor-system specificity
    comparisons, geographic replication, polygenic risk-score adjustment,
    and questionnaire-domain decomposition. \textbf{c}, Molecular and clinical
    characterisation of DK burden using NMR metabolites, routine blood
    biomarkers, plasma proteins, pathway enrichment, exploratory mediation
    analyses, and incident chronic-disease outcomes.}
    \label{fig:1}
\end{figure*}

Here, we investigate these questions in 502,234 UK Biobank participants with baseline touchscreen-questionnaire data and longitudinal health-record linkage \cite{sudlow2015uk,bycroft2018ukb} (Fig.~\ref{fig:1}). We quantified total DK burden and modelled prospective associations with a dementia-related disease composite (NNDs), AD, and vascular dementia (VD). Categorical Cox models and continuous dose--response analyses tested whether risk increased progressively across the observed DK range. A two-year lag analysis evaluated temporal robustness, while participants recruited in Scotland and Wales provided a geographic split-sample replication. We further examined exploratory subgroup heterogeneity and decomposed DK burden across 31 directly answered questionnaire domains. PRS analyses examined genetic correlates of DK burden and tested whether its prospective associations were attenuated, modified, or jointly accumulated with inherited disease susceptibility. Chronic-disease outcomes, nuclear magnetic resonance (NMR) metabolites, routine blood biomarkers, and plasma proteins were used to characterise the wider clinical and molecular context of the signal.

\section*{Results}

\subsection*{Baseline characteristics by DK-response group}

Baseline characteristics were summarised as medians with interquartile ranges
or as numbers with percentages, with percentages calculated using the
non-missing observations for each variable. Differences across the four DK
groups were evaluated using Kruskal--Wallis tests for continuous variables and
Pearson chi-squared tests for categorical variables. Body mass index is
abbreviated as BMI in Table~\ref{tab:baseline}, and disease-history rows
represent prevalence at baseline.

\begin{table}[htbp]
\centering
\caption{Baseline characteristics of UK Biobank participants by number of DK responses.}
\label{tab:baseline}

\resizebox{\textwidth}{!}{%
\begin{tabular}{lccccc}
\hline
\textbf{\# of DK Responses}
& \textbf{0--1}
& \textbf{2--4}
& \textbf{5--7}
& \textbf{$>7$}
& \textbf{\textit{P} Value} \\
\hline
Participants, \(n\)
& 191,536 & 170,928 & 71,546 & 68,224 & \\

Age (years)
& 56.00 [49.00, 62.00]
& 58.00 [51.00, 63.00]
& 59.00 [52.00, 64.00]
& 60.00 [52.00, 64.00]
& $<0.001$ \\

Sex, \(n\) (\%)
& & & & & $<0.001$ \\

\quad Female
& 97,918 (51.1\%)
& 92,331 (54.0\%)
& 42,030 (58.7\%)
& 40,942 (60.0\%)
& \\

\quad Male
& 93,618 (48.9\%)
& 78,597 (46.0\%)
& 29,516 (41.3\%)
& 27,282 (40.0\%)
& \\

BMI (kg\,m$^{-2}$)
& 26.50 [24.02, 29.51]
& 26.72 [24.13, 29.86]
& 26.96 [24.24, 30.23]
& 27.35 [24.49, 30.84]
& $<0.001$ \\

Ethnicity, \(n\) (\%)
& & & & & $<0.001$ \\

\quad White
& 183,483 (95.8\%)
& 163,489 (95.6\%)
& 67,333 (94.1\%)
& 60,028 (88.0\%)
& \\

\quad Non-White
& 8,053 (4.2\%)
& 7,439 (4.4\%)
& 4,213 (5.9\%)
& 8,196 (12.0\%)
& \\

Townsend Deprivation Index
& $-2.41$ [$-3.79$, $-0.05$]
& $-2.21$ [$-3.68$, 0.36]
& $-1.88$ [$-3.49$, 0.99]
& $-1.08$ [$-3.10$, 2.18]
& $<0.001$ \\

Education, \(n\) (\%)
& & & & & $<0.001$ \\

\quad Degree
& 25,975 (13.6\%)
& 19,078 (11.2\%)
& 6,555 (9.2\%)
& 5,234 (7.7\%)
& \\

\quad No Degree
& 165,561 (86.4\%)
& 151,850 (88.8\%)
& 64,991 (90.8\%)
& 62,990 (92.3\%)
& \\

Income, \(n\) (\%)
& & & & & $<0.001$ \\

\quad Greater than 100,000
& 13,600 (7.7\%)
& 7,015 (4.7\%)
& 1,609 (2.9\%)
& 693 (1.5\%)
& \\

\quad 52,000 to 100,000
& 44,569 (25.3\%)
& 28,974 (19.6\%)
& 8,331 (14.8\%)
& 4,341 (9.7\%)
& \\

\quad 31,000 to 51,999
& 48,798 (27.7\%)
& 39,286 (26.6\%)
& 13,739 (24.4\%)
& 8,889 (19.8\%)
& \\

\quad 18,000 to 30,999
& 40,439 (22.9\%)
& 39,054 (26.4\%)
& 15,887 (28.2\%)
& 12,737 (28.4\%)
& \\

\quad Less than 18,000
& 28,802 (16.3\%)
& 33,449 (22.6\%)
& 16,694 (29.7\%)
& 18,213 (40.6\%)
& \\

Smoking Status, \(n\) (\%)
& & & & & $<0.001$ \\

\quad Non-Smoker
& 107,712 (56.2\%)
& 91,455 (53.5\%)
& 37,069 (51.8\%)
& 35,086 (51.4\%)
& \\

\quad Smoker
& 83,824 (43.8\%)
& 79,473 (46.5\%)
& 34,477 (48.2\%)
& 33,138 (48.6\%)
& \\

Physical Activity, \(n\) (\%)
& & & & & $<0.001$ \\

\quad High
& 77,255 (42.4\%)
& 56,281 (41.2\%)
& 17,105 (40.5\%)
& 9,089 (37.8\%)
& \\

\quad Moderate
& 33,702 (18.5\%)
& 23,972 (17.6\%)
& 7,127 (16.9\%)
& 3,927 (16.3\%)
& \\

\quad Low
& 71,375 (39.1\%)
& 56,202 (41.2\%)
& 17,990 (42.6\%)
& 11,055 (45.9\%)
& \\

Sleep Duration, \(n\) (\%)
& & & & & $<0.001$ \\

\quad Long
& 2,704 (1.4\%)
& 3,078 (1.8\%)
& 1,613 (2.3\%)
& 1,846 (2.7\%)
& \\

\quad Normal
& 145,700 (76.4\%)
& 126,225 (73.8\%)
& 50,275 (70.3\%)
& 43,399 (63.6\%)
& \\

\quad Short
& 42,248 (22.2\%)
& 41,621 (24.4\%)
& 19,657 (27.5\%)
& 22,979 (33.7\%)
& \\

Alcohol Consumption (g per week)
& 82.00 [24.00, 160.00]
& 72.00 [12.00, 152.00]
& 68.00 [0.00, 144.00]
& 48.00 [0.00, 120.00]
& $<0.001$ \\

Disease History, \(n\) (\%)
& & & & & \\

\quad Hypertension
& 46,379 (24.2\%)
& 46,664 (27.3\%)
& 20,783 (29.1\%)
& 20,769 (30.4\%)
& $<0.001$ \\

\quad Diabetes
& 8,132 (4.2\%)
& 8,708 (5.1\%)
& 4,370 (6.1\%)
& 5,101 (7.5\%)
& $<0.001$ \\

\quad Dyslipidemia
& 24,992 (13.0\%)
& 26,034 (15.2\%)
& 11,860 (16.6\%)
& 12,548 (18.4\%)
& $<0.001$ \\

\quad Depression
& 13,713 (7.2\%)
& 14,001 (8.2\%)
& 7,045 (9.8\%)
& 7,536 (11.0\%)
& $<0.001$ \\

\quad Cardiovascular Disease
& 8,579 (4.5\%)
& 9,672 (5.7\%)
& 4,692 (6.6\%)
& 5,158 (7.6\%)
& $<0.001$ \\
\hline
\end{tabular}%
}

\end{table}

Among 502,234 participants, 191,536 (38.1\%) had 0--1 DK responses, 170,928
(34.0\%) had 2--4 responses, 71,546 (14.2\%) had 5--7 responses, and 68,224
(13.6\%) had more than seven responses. Baseline characteristics differed
across all four DK groups (all omnibus \(P<0.001\); Table~\ref{tab:baseline}).
Higher DK burden was accompanied by progressive differences in demographic,
socioeconomic, lifestyle, and clinical profiles.

Participants with more than seven DK responses were older than those with
0--1 responses (median age, 60 versus 56 years), and the proportion of women
increased from 51.1\% to 60.0\% across these groups. Median BMI increased from
26.50 to 27.35~kg\,m\(^{-2}\). The proportion of non-White participants was
4.2\% in the 0--1 group and 12.0\% in the \(>7\) group. Socioeconomic
differences were similarly graded. The median Townsend Deprivation Index
increased from \(-2.41\) to \(-1.08\), while the proportion holding a degree
decreased from 13.6\% to 7.7\%. Among participants with available income data,
the proportion reporting an annual household income above 100,000 decreased
from 7.7\% to 1.5\%. The proportion reporting an income below 18,000 increased
from 16.3\% to 40.6\%.

Lifestyle characteristics also varied across DK groups. From the lowest to
the highest group, the proportion of smokers increased from 43.8\% to 48.6\%,
while the proportion reporting high physical activity decreased from 42.4\%
to 37.8\%. Short sleep was reported by 22.2\% of participants with 0--1 DK
responses and by 33.7\% of those with more than seven responses. Long sleep
also became more frequent, increasing from 1.4\% to 2.7\%. Conversely, median
weekly alcohol consumption decreased from 82 to 48~g across the same groups.

Prevalent cardiometabolic and mental-health conditions were more common at
higher levels of DK burden. Hypertension increased from 24.2\% in the 0--1
group to 30.4\% in the \(>7\) group, and diabetes increased from 4.2\% to
7.5\%. Corresponding increases were observed for dyslipidaemia (13.0\% to
18.4\%), depression (7.2\% to 11.0\%), and cardiovascular disease (4.5\% to
7.6\%). These systematic differences provided the rationale for sequential
adjustment for demographic, socioeconomic, behavioural, and clinical
characteristics in the prospective analyses.

\subsection*{DK burden and incident dementia-related outcomes}

\begin{figure}[htbp]
    \centering
    \includegraphics[width=\linewidth]{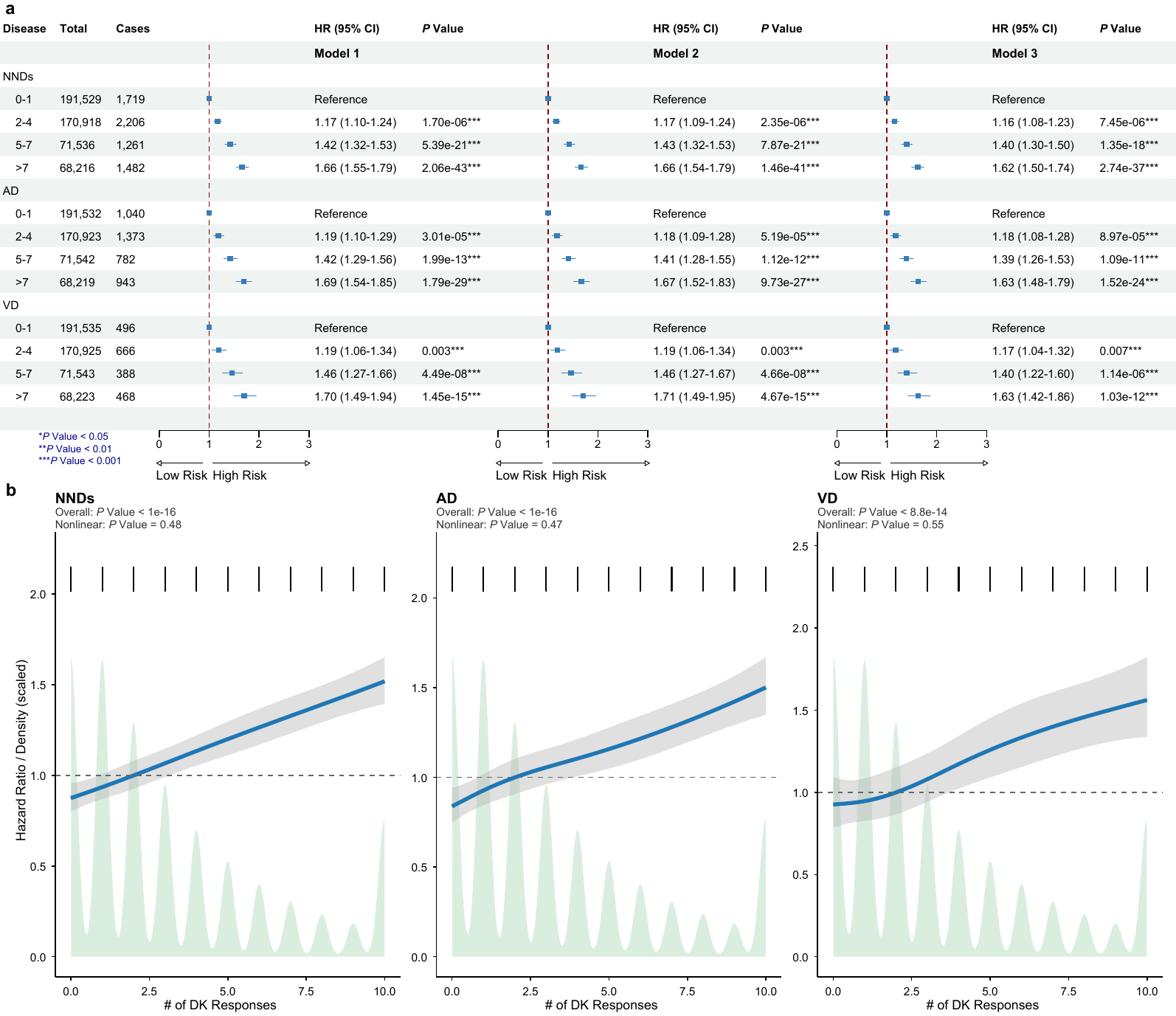}
    \caption{\textbf{Prospective dose--response associations between DK
    burden and incident dementia-related outcomes.} \textbf{a}, Hazard ratios
    and 95\% confidence intervals for NNDs, AD, and VD across four DK response
    groups. Participants reporting 0--1 responses were the reference. Model~1
    included demographic and socioeconomic covariates; Model~2 additionally
    included smoking, physical activity, sleep duration, and alcohol
    consumption; Model~3 further included prevalent cardiovascular disease,
    depression, dyslipidaemia, diabetes, and hypertension. \textbf{b}, Fully
    adjusted restricted cubic spline estimates for continuous DK count. Counts
    above ten were truncated at ten, and the median count of two was the
    reference. Solid lines show estimated hazard ratios, shaded bands show
    95\% confidence intervals, horizontal dashed lines mark a hazard ratio of
    1, pale areas show the scaled DK-count distributions, and upper marks show
    event locations. Overall and nonlinearity \(P\) values are displayed in
    each panel.}
    \label{fig:2}
\end{figure}

During follow-up, 6,668 incident NND cases, 4,138 AD cases, and 2,018 VD
cases were recorded. Incident outcomes and the
time-to-event framework are described in
\hyperref[method:outcomes]{Methods: Ascertainment of incident disease outcomes}
and
\hyperref[method:primary]{Methods: Primary Cox proportional hazards analyses}.
Across all three progressively adjusted models, a higher number of DK responses
was associated with a graded increase in the incidence of each outcome
(Fig.~\ref{fig:2}a; Supplementary Table S4). Because the estimates were
broadly stable across models, the fully adjusted Model~3 estimates are
reported below.

Compared with participants reporting 0--1 DK responses, the adjusted hazard
ratios (HRs) for NNDs were 1.16 (95\% confidence interval [CI],
1.08--1.23) for 2--4 responses, 1.40 (95\% CI, 1.30--1.50) for 5--7
responses, and 1.62 (95\% CI, 1.50--1.74) for more than seven responses.
The corresponding HRs for AD were 1.18 (95\% CI, 1.08--1.28), 1.39
(95\% CI, 1.26--1.53), and 1.63 (95\% CI, 1.48--1.79), respectively.
A similar gradient was observed for VD, with HRs of 1.17 (95\% CI,
1.04--1.32), 1.40 (95\% CI, 1.22--1.60), and 1.63 (95\% CI,
1.42--1.86), respectively. Thus, the association was not confined to the
highest DK group but increased progressively across the prespecified response
categories.

When DK burden was analysed as a continuous exposure using restricted cubic
splines
(\hyperref[method:rcs]{Methods: Continuous dose--response analyses}), the
results were consistent with the categorical analyses (Fig.~\ref{fig:2}b).
There was strong evidence of an overall association for NNDs and AD
(\(P_{\mathrm{overall}}<1\times10^{-16}\) for both) and for VD
(\(P_{\mathrm{overall}}<8.8\times10^{-14}\)). However, there was no
statistical evidence of departure from linearity
(\(P_{\mathrm{nonlinear}}=0.48\), 0.47, and 0.55 for NNDs, AD, and VD,
respectively). The fitted curves therefore indicated an approximately linear
increase in disease incidence across the observed range of DK counts, with
wider confidence intervals towards the sparsely populated upper end of the
distribution.
Curve coordinates and overall and nonlinearity tests are reported in
Supplementary Table S5.

To reduce the possibility that the principal associations were driven by
disease already close to clinical recognition at baseline, we repeated the
fully adjusted analysis after excluding diagnoses occurring within the first
730 days of follow-up
(\hyperref[method:negative-controls]{Methods: Motor-system comparisons and two-year
lag analyses}).
After this exclusion, 6,600 NND, 4,100 AD, and 1,997 VD events remained.
The graded associations were essentially unchanged
(Supplementary Fig.~S3). Compared with the 0--1 response group,
participants with more than seven DK responses had an HR of 1.62
(95\% CI, 1.50--1.74) for NNDs, 1.63 (95\% CI, 1.49--1.79) for AD,
and 1.64 (95\% CI, 1.43--1.88) for VD. Intermediate DK groups also
retained progressively elevated HRs. Complete lagged estimates are reported in
Supplementary Table S6 and support the robustness of the main gradient to
removal of diagnoses recorded soon after baseline.

We additionally examined motor-system degenerative outcomes using the same Cox modelling framework (\hyperref[method:negative-controls]{Methods: Motor-system comparisons and two-year lag analyses}). Associations were smaller than those for dementia-related outcomes but were not absent: in Model~3, the HRs comparing more than seven with 0--1 DK responses were 1.24 for motor neurone disease and 1.24 for Parkinson's disease, whereas the amyotrophic lateral sclerosis estimate was less precise and compatible with no association (HR, 1.22; 95\% CI, 0.95--1.58). Thus, higher DK burden was associated with selected motor-system outcomes, although less strongly than with dementia-related outcomes (Supplementary Fig.~S1 and Supplementary Table S7).

\subsection*{Geographic replication and subgroup heterogeneity}

We next assessed whether the association between DK response burden and
dementia-related outcomes was reproducible across geographically distinct
subsets of the UK Biobank
(\hyperref[method:geographic-validation]{Methods: Geographic split-sample
replication}). Participants recruited in England were used as the development
sample, while those recruited in Scotland and Wales formed a geographically
separate internal replication sample. After applying a common administrative
censoring date and censoring at death when it preceded the outcome, the development
sample included 445,566 participants and 5,470 NND events, 445,583
participants and 3,384 AD events, and 445,591 participants and 1,658 VD
events. The corresponding replication samples included 56,633 participants
and 713 NND events, 56,633 participants and 426 AD events, and 56,635
participants and 237 VD events.
Sample construction and outcome counts are detailed in Supplementary Table S8,
with complete region-specific estimates and interaction tests in Supplementary
Tables S9 and S10.

The risk gradients observed in England were directionally reproduced in
Scotland and Wales, although estimates in the smaller replication sample were
less precise (Table~\ref{tab:geographic-validation}). For NNDs, the fully
adjusted HR for more than seven versus 0--1 DK responses was 1.75
(95\% CI, 1.62--1.90) in England and 1.53 (95\% CI, 1.21--1.93) in
Scotland and Wales. The corresponding HRs were 1.80 (95\% CI,
1.62--1.99) and 1.36 (95\% CI, 1.01--1.83) for AD, and 1.77
(95\% CI, 1.53--2.05) and 1.67 (95\% CI, 1.11--2.52) for VD.
The 5--7 response group was also associated with elevated NND and VD
incidence in the replication sample, whereas estimates for the lower DK
categories were weaker and, for several comparisons, had confidence intervals
that included one.

Table~\ref{tab:geographic-validation} reports the fully adjusted Model~3
estimates separately for the England development sample and the
Scotland-and-Wales replication sample. Participants with 0--1 DK responses were the reference group. Interaction \(P\) values evaluate the overall interaction between the four-level DK variable and geographic sample. Evidence of geographic heterogeneity was detected for AD (\(P_{\mathrm{interaction}}=0.028\)) but not for NNDs or VD.

Together, these estimates showed that the association for the highest DK group was directionally reproduced in the geographically distinct replication
sample, with wider confidence intervals reflecting its smaller number of
participants and events.

We additionally examined the association with NNDs within prespecified
demographic, socioeconomic, lifestyle, and baseline health strata
(\hyperref[method:subgroups]{Methods: Subgroup and interaction analyses}).
The association for more than seven versus 0--1 DK responses was positive
across most strata, including women (HR, 1.68; 95\% CI, 1.51--1.87), men
(HR, 1.55; 95\% CI, 1.39--1.72), participants with and without prevalent
cardiovascular disease, and participants with and without hypertension,
diabetes, dyslipidaemia, or depression (Supplementary Table S12). Formal
Exploratory DK-by-covariate interaction tests provided nominal evidence of effect heterogeneity by household
income (\(P_{\mathrm{interaction}}=7.8\times10^{-6}\)), smoking status
(\(P_{\mathrm{interaction}}=0.006\)), educational attainment
(\(P_{\mathrm{interaction}}=0.028\)), and ethnicity
(\(P_{\mathrm{interaction}}=0.029\); Supplementary Table S13). For the
highest DK group, the HR ranged from 3.75 (95\% CI, 1.78--7.88) among
participants with household income above 100,000 to 1.30 (95\% CI,
1.15--1.47) among those with household income below 18,000. The corresponding
HR was higher among non-smokers than smokers (1.85 versus 1.44), among
degree-holders than participants without a degree (2.08 versus 1.59), and
among White than non-White participants (1.66 versus 1.21). There was no
evidence of interaction by age, sex, BMI, Townsend deprivation, alcohol
consumption, physical activity, sleep duration, or baseline health conditions
(all \(P_{\mathrm{interaction}}>0.05\)). Thus, the high-DK association was observed across a broad range of participant characteristics, with possible variation across selected socioeconomic and behavioural strata.

\begin{table*}[htbp]
\centering
\caption{Geographic split-sample replication of multivariable-adjusted associations
between DK response groups and incident dementia-related outcomes in the
England development sample and the Scotland-and-Wales replication sample.}
\label{tab:geographic-validation}
\small
\begin{tabular}{llccc}
\toprule
\textbf{Outcome} &
\textbf{DK response group} &
\textbf{England, HR (95\% CI)} &
\textbf{Scotland and Wales, HR (95\% CI)} &
\textbf{\textit{P} for interaction} \\
\midrule
NNDs
& 2--4 & 1.21 (1.13--1.30) & 1.05 (0.87--1.27) & 0.464 \\
& 5--7 & 1.50 (1.38--1.62) & 1.37 (1.10--1.71) & \\
& $>7$ & 1.75 (1.62--1.90) & 1.53 (1.21--1.93) & \\
\midrule
AD
& 2--4 & 1.27 (1.16--1.39) & 0.87 (0.68--1.11) & 0.028 \\
& 5--7 & 1.51 (1.36--1.68) & 1.24 (0.93--1.65) & \\
& $>7$ & 1.80 (1.62--1.99) & 1.36 (1.01--1.83) & \\
\midrule
VD
& 2--4 & 1.20 (1.05--1.37) & 1.18 (0.85--1.65) & 0.859 \\
& 5--7 & 1.51 (1.30--1.75) & 1.66 (1.13--2.44) & \\
& $>7$ & 1.77 (1.53--2.05) & 1.67 (1.11--2.52) & \\
\bottomrule
\end{tabular}

\end{table*}

\subsection*{DK burden remained associated after PRS adjustment}
\label{results:prs}

To determine whether DK burden primarily reflected measured polygenic
susceptibility, we first examined associations between standardised DK count
and 39 Standard and 51 Enhanced PRSs using covariate-adjusted linear
regression (\hyperref[method:prs]{Methods: Polygenic risk-score analyses}).
Of these, 22 Standard PRSs and 33 Enhanced PRSs were associated with DK burden
after Benjamini--Hochberg false discovery rate (FDR) correction. However, the
variance attributable to any individual PRS was small: the largest partial
\(R^{2}\) was \(7.99\times10^{-4}\) (0.080\%) in the Standard set and
\(2.29\times10^{-4}\) (0.023\%) in the Enhanced set. Among the Standard
scores most relevant to dementia-related and vascular susceptibility, higher
PRSs for ischaemic stroke (\(\beta=0.0080\) standard deviations of DK burden
per 1-standard-deviation increase in PRS; 95\% CI, 0.0053--0.0107;
FDR-adjusted \(P=4.8\times10^{-8}\)), hypertension (\(\beta=0.0069\);
95\% CI, 0.0042--0.0096; FDR-adjusted \(P=2.6\times10^{-6}\)), and
cardiovascular disease (\(\beta=0.0042\); 95\% CI, 0.0015--0.0068;
FDR-adjusted \(P=0.007\)) were associated with greater DK burden. By
contrast, the association with the AD PRS did not pass FDR correction
(\(\beta=0.0029\); 95\% CI, 0.0003--0.0056; FDR-adjusted \(P=0.053\)).
Complete associations for both PRS sets are reported in Supplementary
Table S24.

We next evaluated eight prespecified disease--PRS pairs in both the Standard
and Enhanced sets, producing 16 targeted Cox analyses. The PRS main effect was
FDR-significant in 14 of the 16 models. As expected, the Standard AD PRS was
strongly associated with incident NNDs (HR per 1-standard-deviation increase,
1.81; 95\% CI, 1.78--1.85) and AD (HR, 1.98; 95\% CI, 1.93--2.03). For VD,
the Standard cardiovascular-disease and ischaemic-stroke PRSs were associated
with HRs of 1.11 (95\% CI, 1.06--1.16) and 1.13 (95\% CI, 1.08--1.18),
respectively (Supplementary Table S25).

Adjustment for the corresponding PRS produced little change in the association
between continuous DK burden and disease risk. Across all 16 models, the
change in the DK log-HR ranged from a 1.60\% strengthening to a 2.84\%
attenuation, and the adjusted DK association remained FDR-significant in every
model. After adjustment for the Standard AD PRS, each 1-standard-deviation
increase in DK count was associated with incident NNDs (HR, 1.145; 95\% CI,
1.121--1.168; FDR-adjusted \(P=1.5\times10^{-37}\)) and AD (HR, 1.142;
95\% CI, 1.113--1.172; FDR-adjusted \(P=7.3\times10^{-23}\)). After
adjustment for the Standard cardiovascular-disease PRS, the corresponding
association with VD was HR 1.158 (95\% CI, 1.117--1.200; FDR-adjusted
\(P=1.9\times10^{-15}\)).

None of the 16 multiplicative DK-by-PRS interaction terms passed FDR
correction (all FDR-adjusted \(P\geq0.947\)). Nevertheless, participants with
both more than seven DK responses and a PRS in the highest quintile had
substantially higher risks than those with no more than seven DK responses and
a PRS in the lowest quintile. The joint high-DK/high-PRS HR was 6.55 (95\% CI,
5.81--7.39; FDR-adjusted \(P=2.3\times10^{-204}\)) for NNDs using the
Standard AD PRS, 8.49 (95\% CI, 7.27--9.93; FDR-adjusted
\(P=6.9\times10^{-159}\)) for AD using the same PRS, and 1.85 (95\% CI,
1.47--2.33; FDR-adjusted \(P=1.9\times10^{-7}\)) for VD using the Standard
cardiovascular-disease PRS. Complete adjustment,
interaction, joint-risk, model-comparison, and diagnostic results are provided
in Supplementary Table S25. Overall, DK burden remained associated with all
three dementia-related outcomes after PRS adjustment. Participants with both
high DK burden and high PRS had especially elevated risks, without evidence
of multiplicative interaction.

\subsection*{Questionnaire domains showed distinct associations}
\label{results:questionnaire-domains}

\begin{figure*}[htbp]
    \centering
    \includegraphics[width=\textwidth]{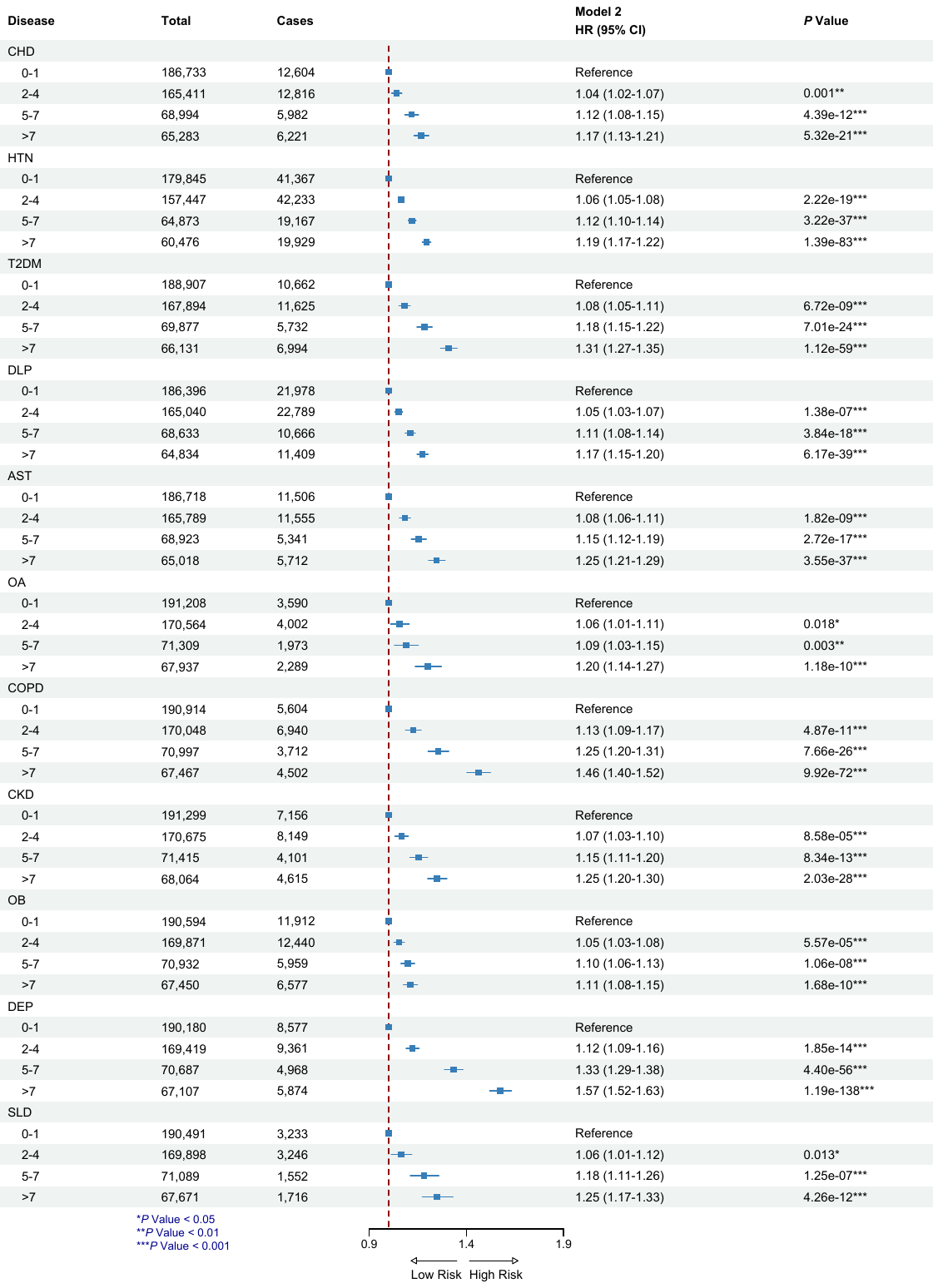}
    \caption{\textbf{Questionnaire-domain-specific DK responses and risks of
    dementia and chronic diseases.} The heatmap presents hazard ratios for AD,
    VD, chronic ischaemic heart disease, hypertension, type 2 diabetes,
    disorders of lipoprotein metabolism, asthma, and polyarthrosis. Columns
    represent UK Biobank touchscreen domains, and each cell represents the
    hazard ratio per additional DK response within that domain. All estimable
    domain counts were entered simultaneously into the fully adjusted Cox
    model, so each estimate was conditional on DK responses in the other
    domains. Red and blue cells indicate hazard ratios above and below 1,
    respectively, with colour intensity proportional to association magnitude.
    Only associations passing Benjamini--Hochberg FDR correction are coloured;
    neutral cells were not FDR significant or were not estimable. Asterisks
    indicate FDR-adjusted \(P<0.05\), \(P<0.01\), and \(P<0.001\).}
    \label{fig:6}
\end{figure*}

To determine whether the association between DK responses and subsequent disease risk was distributed uniformly across the touchscreen questionnaire or concentrated within specific types of information, we decomposed total DK burden into questionnaire-domain-specific counts. All estimable domain counts were entered simultaneously into the fully adjusted Cox model, thereby estimating each domain independently of DK responses in the others. Statistical significance was evaluated using Benjamini--Hochberg FDR correction across estimable disease--domain associations (Fig.~\ref{fig:6}; \hyperref[method:questionnaire-domains]{Methods: Questionnaire-domain-specific analyses}). Complete estimates are reported in Supplementary Table S14.

The domain-specific analysis revealed a structured rather than uniform pattern. DK responses in the \textit{Household} domain showed the strongest associations with both AD (HR per additional domain-specific DK response, 1.51; 95\% CI, 1.34--1.69; FDR-adjusted \(P<0.001\)) and VD (HR, 1.50; 95\% CI, 1.27--1.77; FDR-adjusted \(P<0.001\)). DK responses concerning \textit{Medical Conditions} were also associated with both outcomes. Additional AD associations involved the \textit{Medication}, \textit{Cancer Screening}, and \textit{Family History} domains, whereas VD was additionally associated with \textit{Chest Pain} and \textit{Electronic Device Use} (Fig.~\ref{fig:6}). Mutual adjustment across domains indicated that these patterns were not attributable solely to a general tendency to provide uncertain responses.

The cross-disease comparison provided further context for these dementia-related patterns. DK responses in the \textit{Medical Conditions} and \textit{Chest Pain} domains were consistently associated with several chronic disease outcomes, including chronic ischaemic heart disease, hypertension, type 2 diabetes, disorders of lipoprotein metabolism, asthma, and polyarthrosis. In contrast, the \textit{Household} domain produced the most prominent estimates in the AD and VD rows, despite containing information that was not itself a direct clinical assessment of dementia. Thus, Figure~\ref{fig:6} indicates that DK responses do not constitute a single undifferentiated form of missingness: their prospective associations vary according to both the information being queried and the subsequent disease outcome.

Together, these findings localise the aggregate signal: uncertainty when recalling or reporting ordinary household and personal information contributes information not represented fully by conventional clinical covariates, while medical and symptom-related domains connect DK burden to broader health vulnerability.

\subsection*{Metabolic and biochemical correlates of DK burden}
\label{results:metabolic-biochemical}

\begin{figure*}[htbp]
    \centering
    \includegraphics[width=0.98\textwidth]{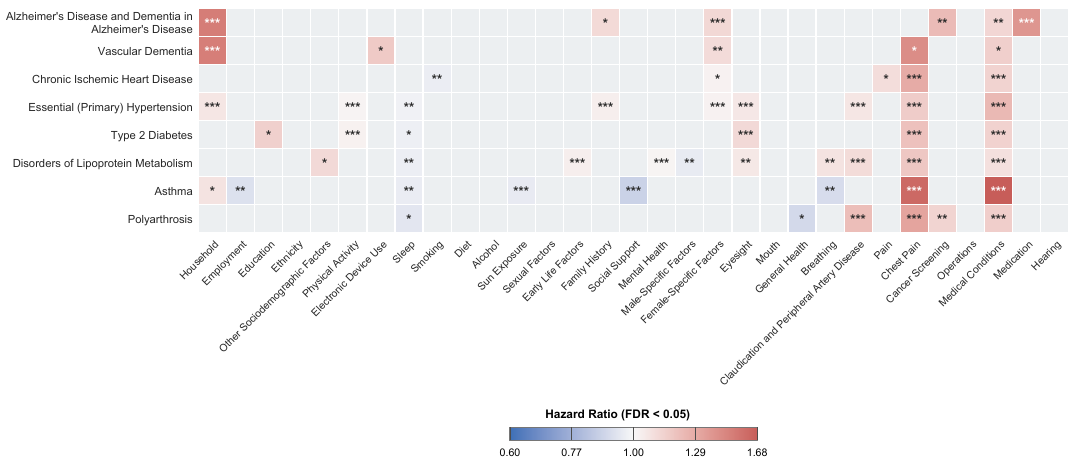}
    \caption{\textbf{Metabolic and biochemical correlates of DK burden and
    exploratory mediation estimates.} \textbf{a}, Associations between
    continuous DK count and 251 standardised NMR-derived metabolic traits,
    grouped by biochemical category. The vertical axis represents
    \(-\log_{10}\) of the FDR-adjusted \(P\) value, and point colour represents
    the direction and magnitude of the regression coefficient. \textbf{b},
    Corresponding associations with 37 routinely measured or derived blood
    biomarkers, grouped by physiological system. \textbf{c}, Regression
    coefficients and 95\% confidence intervals for selected FDR-significant
    NMR traits and blood biomarkers. Coefficients represent the
    standard-deviation difference per additional DK response. \textbf{d},
    FDR-significant model-based TNIE estimates for DK count, candidate circulating
    markers, and incident NNDs, AD, or VD. Point colour represents the log
    indirect-effect hazard ratio, and point size represents
    \(-\log_{10}\) of the FDR-adjusted \(P\) value. Association and
    mediation models used the fully adjusted covariate set.}
    \label{fig:3}
\end{figure*}

We next examined whether DK response burden was accompanied by measurable differences in circulating metabolic and biochemical phenotypes. Associations with NMR metabolites and routinely measured blood biomarkers were evaluated using covariate-adjusted linear regression, with each molecular outcome standardised before modelling. Coefficients therefore represent the standard-deviation difference in a marker per additional DK response. Multiple testing was controlled using the Benjamini--Hochberg FDR procedure (Fig.~\ref{fig:3}; \hyperref[method:nmr-metabolomics]{Methods: NMR metabolomic analyses}; \hyperref[method:blood-biochemistry]{Methods: Blood-biochemistry analyses}). Complete association estimates are reported in Supplementary Tables S16 and S18.

Among 251 NMR-derived metabolic traits assessed in 212,853 participants, 186 were associated with DK count at FDR-adjusted \(P<0.05\) (Fig.~\ref{fig:3}a). Although individual effect sizes were modest, the associations converged on fatty-acid composition and high-density lipoprotein (HDL)-related lipid transport. Higher DK counts were associated with a lower ratio of docosahexaenoic acid (DHA) to total fatty acids (\(\beta=-0.0081\) standard deviations per additional response; 95\% CI, \(-0.0090\) to \(-0.0071\); FDR-adjusted \(P=1.5\times10^{-61}\)) and lower absolute DHA concentrations (\(\beta=-0.0076\); 95\% CI, \(-0.0086\) to \(-0.0067\); FDR-adjusted \(P=4.7\times10^{-56}\)). Consistent inverse associations involved omega-3 fatty acids and several HDL measures, whereas higher DK counts were associated with a higher omega-6-to-omega-3 ratio and glycoprotein acetyl levels. The resulting pattern spanned fatty-acid balance, HDL composition, and systemic inflammatory activity.

These metabolic features overlap with circulating signatures previously associated with dementia. DHA is highly enriched in central-nervous-system membrane phospholipids, and prospective UK Biobank analyses have associated higher circulating total omega-3 and DHA measures with lower subsequent risks of all-cause dementia and AD \cite{sala2023omega3}. Independent NMR analyses in UK Biobank have further linked fatty-acid measures, lipoprotein subclasses including small-HDL components, ketone bodies, glucose, and valine with incident dementia, AD, or VD \cite{huang2023metabolites}. The lower DHA and omega-3 measures and altered HDL-related profile observed with increasing DK burden therefore place questionnaire response uncertainty within a lipid and metabolic axis that has independently been associated with subsequent dementia-related disease.

Of the 37 routinely measured or derived blood biomarkers, 26 were associated with DK count after FDR correction (Fig.~\ref{fig:3}b,c). Higher DK burden was associated with lower vitamin D (\(\beta=-0.0101\); 95\% CI, \(-0.0108\) to \(-0.0095\); FDR-adjusted \(P=1.4\times10^{-201}\)), a lower creatinine-to-cystatin C ratio (\(\beta=-0.0087\); 95\% CI, \(-0.0093\) to \(-0.0082\); FDR-adjusted \(P=7.2\times10^{-207}\)), and lower HDL cholesterol (\(\beta=-0.0041\); 95\% CI, \(-0.0047\) to \(-0.0035\); FDR-adjusted \(P=1.0\times10^{-42}\)). Positive associations involved cystatin C, glycated haemoglobin, C-reactive protein, alkaline phosphatase, and inflammatory cell-count indices. These directions also parallel prospective dementia-associated laboratory profiles. In more than 400,000 UK Biobank participants, incident dementia was associated with lower vitamin D, renal abnormalities indexed by cystatin C, higher glucose and glycated haemoglobin, and dysregulation of C-reactive protein \cite{he2022serum}. Thus, the DK-associated biochemical profile encompasses several systemic processes---nutritional status, renal function, glycaemic regulation, lipid metabolism, and inflammation---that have also been linked prospectively to dementia risk.

We next conducted exploratory mediation analyses with candidate NMR metabolites and blood biomarkers (Fig.~\ref{fig:3}d; \hyperref[method:indirect-effects]{Methods: Exploratory molecular mediation analyses}). Among 558 outcome--metabolite tests, 76 passed outcome-specific FDR correction: 63 for NNDs, one for AD, and 12 for VD. Figure~\ref{fig:3}d displays representative signals, including valine, 3-hydroxybutyrate, and mean very-low-density lipoprotein particle diameter; valine was identified across all three outcomes. Among 78 outcome--blood-marker tests, 21 passed FDR correction, comprising 12 for NNDs, one for AD, and eight for VD. These included glycated haemoglobin, the creatinine-to-cystatin C ratio, vitamin D, C-reactive protein, and inflammatory cell-count indices. Complete total natural indirect effect (TNIE) estimates are reported in Supplementary Tables S17 and S19.

Overall, higher DK burden was associated with a multisystem molecular profile involving altered fatty-acid and lipoprotein metabolism, amino-acid and ketone-body metabolism, nutritional status, glycaemic regulation, renal function, and systemic inflammation. The overlap of these signals with circulating profiles reported before dementia onset provides a biological connection between cumulative questionnaire response uncertainty and the systemic metabolic and inflammatory milieu associated with subsequent dementia-related disease.

\subsection*{Proteomic correlates and pathway enrichment}
\label{results:proteomics}

\begin{figure*}[htbp]
    \centering
    \includegraphics[width=\textwidth]{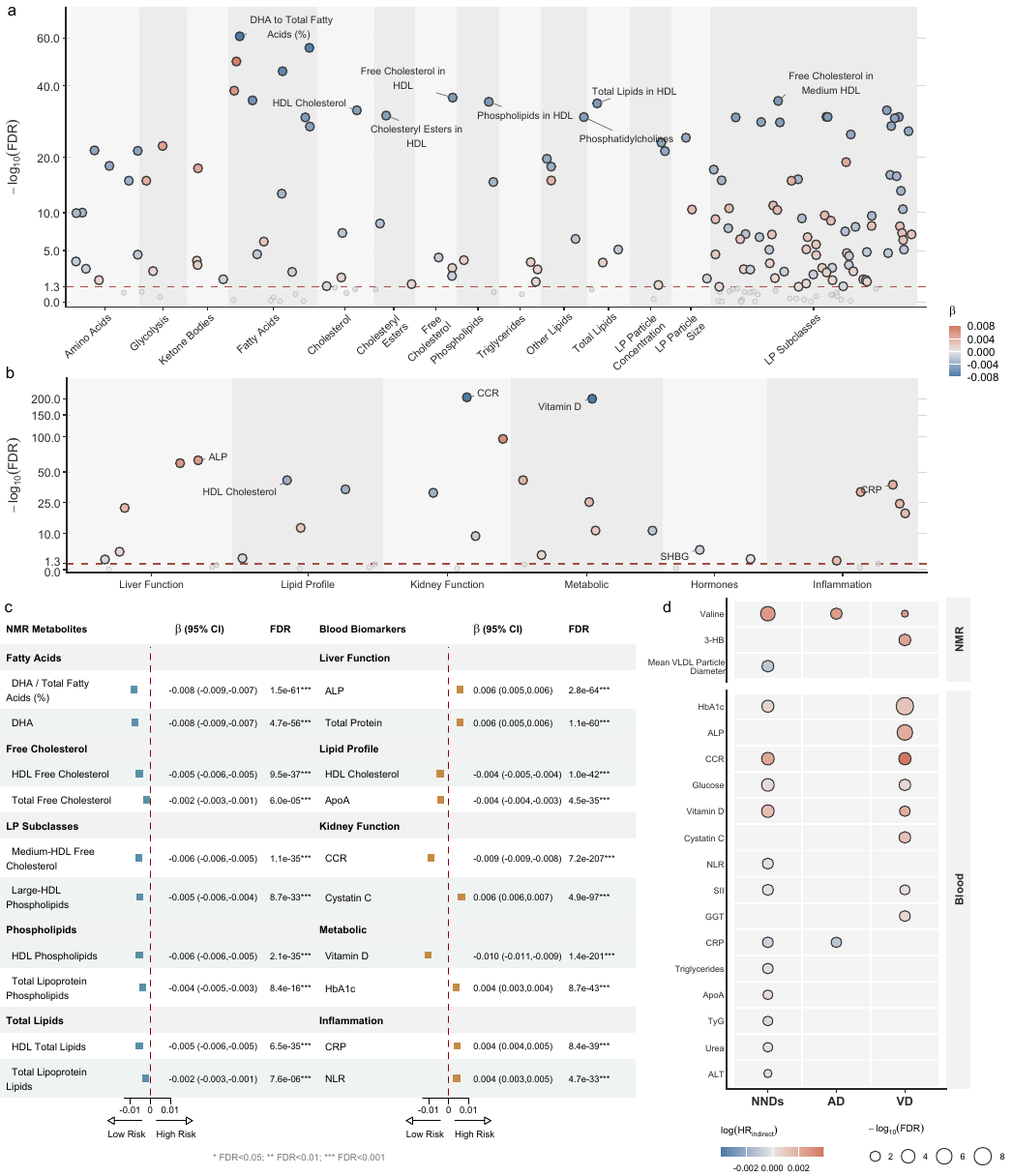}
    \caption{\textbf{Plasma-protein associations with DK burden, pathway
    enrichment, and exploratory mediation estimates.} \textbf{a}, Volcano plot
    of associations between continuous DK count and 2,923 plasma proteins.
    The horizontal axis shows the coefficient per additional DK response, and
    the vertical axis shows \(-\log_{10}\) of the FDR-adjusted \(P\) value.
    Red and blue points indicate significant positive and inverse associations;
    grey points did not pass FDR correction. \textbf{b}, Selected significant
    pathways from ranked gene-set enrichment analysis. Bars show normalised
    enrichment scores and colours distinguish pathway collections.
    \textbf{c}, Selected Gene Ontology biological processes from
    over-representation analysis of FDR-significant proteins. The horizontal
    axis shows fold enrichment, point size shows the number of mapped proteins,
    and colour shows \(-\log_{10}\) of the FDR-adjusted \(P\) value.
    \textbf{d}, Outcome-specific FDR-significant model-based TNIE estimates for DK
    count, candidate proteins, and incident NNDs or VD. Red and blue
    connections indicate positive and inverse associations, respectively. No
    candidate protein passed FDR correction for AD. Protein-association and
    mediation models used the fully adjusted covariate set.}
    \label{fig:4}
\end{figure*}

To further characterise the circulating molecular phenotype, we modelled 2,923 plasma proteins as functions of continuous DK count with adjustment for demographic, socioeconomic, behavioural, and baseline clinical characteristics. The Benjamini--Hochberg FDR procedure was followed by complementary ranked-set and over-representation pathway analyses (Fig.~\ref{fig:4}; \hyperref[method:proteomic-associations]{Methods: Proteomic association analyses}; \hyperref[method:pathway-analysis]{Methods: Proteomic pathway analyses}).

Of the 2,923 proteins examined, 1,363 were associated with DK count at FDR-adjusted \(P<0.05\), including 1,300 positive and 63 inverse associations (Fig.~\ref{fig:4}a). Individual coefficients were small but widespread. The most robust positive associations included plasminogen activator urokinase receptor (\textit{PLAUR}), hepatocyte growth factor (\textit{HGF}), glyoxalase~I (\textit{GLO1}), growth differentiation factor~15 (\textit{GDF15}), oxidized low-density lipoprotein receptor~1 (\textit{OLR1}), oncostatin M (\textit{OSM}), and fatty-acid-binding protein~1 (\textit{FABP1}). Carbonic anhydrase~14 (\textit{CA14}) and oligodendrocyte myelin glycoprotein (\textit{OMG}) were among the strongest inverse associations. Complete protein-level estimates are reported in Supplementary Table S20. Among these signals, \textit{GDF15} provides particularly direct convergence with existing dementia proteomics: an independent prospective analysis of 52,645 UK Biobank participants identified GDF15 among the proteins most consistently associated with future all-cause dementia, AD, and VD \cite{guo2024proteomic}. Its emergence among the strongest DK-associated proteins therefore connects response uncertainty to a circulating protein signature detectable years before dementia diagnosis.

Ranked gene-set enrichment analysis demonstrated that the protein associations were concentrated in biological processes (Fig.~\ref{fig:4}b). The strongest enriched pathway involved tumour necrosis factors binding their physiological receptors (normalised enrichment score [NES], 1.96; FDR-adjusted \(P=0.029\)). Additional enriched processes included neutrophil degranulation (NES, 1.56; FDR-adjusted \(P=0.029\)), T-cell proliferation (NES, 1.55; FDR-adjusted \(P=0.034\)), negative regulation of cell activation, humoral immune responses, innate immune-system activity, and responses to bacterial exposure. Complete enrichment and over-representation results are reported in Supplementary Table S21.

These pathway-level signals also converge with prospective proteomic studies of dementia. In the ARIC cohort, proteome-wide analyses of individuals initially free of dementia identified future dementia-associated networks involving innate immunity, lipid and metabolic signalling, vascular function and haemostasis; the strongest protein-network association involved an immune module enriched for cytokine and myeloid signalling \cite{walker2021proteomic}. The enrichment of TNF-receptor binding, neutrophil activity, T-cell proliferation, and innate and humoral immune responses in the present analysis therefore places the DK-associated proteome within a broader peripheral immune and vascular-metabolic programme already linked to dementia susceptibility.

Over-representation analysis of the FDR-significant proteins produced a consistent pattern (Fig.~\ref{fig:4}c). Enriched Gene Ontology biological processes clustered into three broad themes: immune regulation, host--pathogen interaction, and cellular response. These included positive regulation of immune-system processes, regulation of lymphocyte activation, responses to bacteria and symbionts, regulation of viral processes, negative regulation of protein metabolism, and protein deubiquitination. The analysis used the exploratory code-specified FDR threshold of 0.20, although all terms displayed in Fig.~\ref{fig:4}c had FDR-adjusted \(P<0.05\). Together, the ranked-set and over-representation results identify a coordinated DK-associated protein phenotype characterised by cytokine-receptor signalling, leukocyte activation, host defence, and cellular stress rather than isolated changes in individual circulating proteins.

We next evaluated the 30 most statistically significant DK-associated proteins in exploratory mediation models (Fig.~\ref{fig:4}d; \hyperref[method:protein-indirect-effects]{Methods: Protein mediation analysis}). After outcome-specific FDR correction, \textit{OMG} (indirect-effect HR, 1.0044; 95\% CI, 1.0022--1.0067; FDR-adjusted \(P=0.0038\)) and \textit{GDF15} (HR, 1.0033; 95\% CI, 1.0011--1.0054; FDR-adjusted \(P=0.042\)) were identified for NNDs. For VD, the corresponding proteins were \textit{GDF15} (HR, 1.0078; 95\% CI, 1.0038--1.0118; FDR-adjusted \(P=0.0037\)), mannose receptor C-type~1 (\textit{MRC1}; HR, 1.0073; 95\% CI, 1.0029--1.0118; FDR-adjusted \(P=0.018\)), and regenerating family member~3 alpha (\textit{REG3A}; HR, 1.0054; 95\% CI, 1.0018--1.0090; FDR-adjusted \(P=0.033\)). No candidate protein passed FDR correction for AD. Complete estimates are reported in Supplementary Table S22.

Overall, model-based indirect-effect estimates involved only a small subset of DK-associated proteins, but the recurrence of \textit{GDF15} in both NND and VD models and the broader enrichment of immune-response pathways link the statistical mediation signals to the same systemic stress and immune processes identified in the protein-association analysis.

\subsection*{DK burden and chronic-disease vulnerability}
\label{results:chronic-diseases}

\begin{figure*}[htbp]
    \centering
    \includegraphics[width=\textwidth,height=0.78\textheight,keepaspectratio]{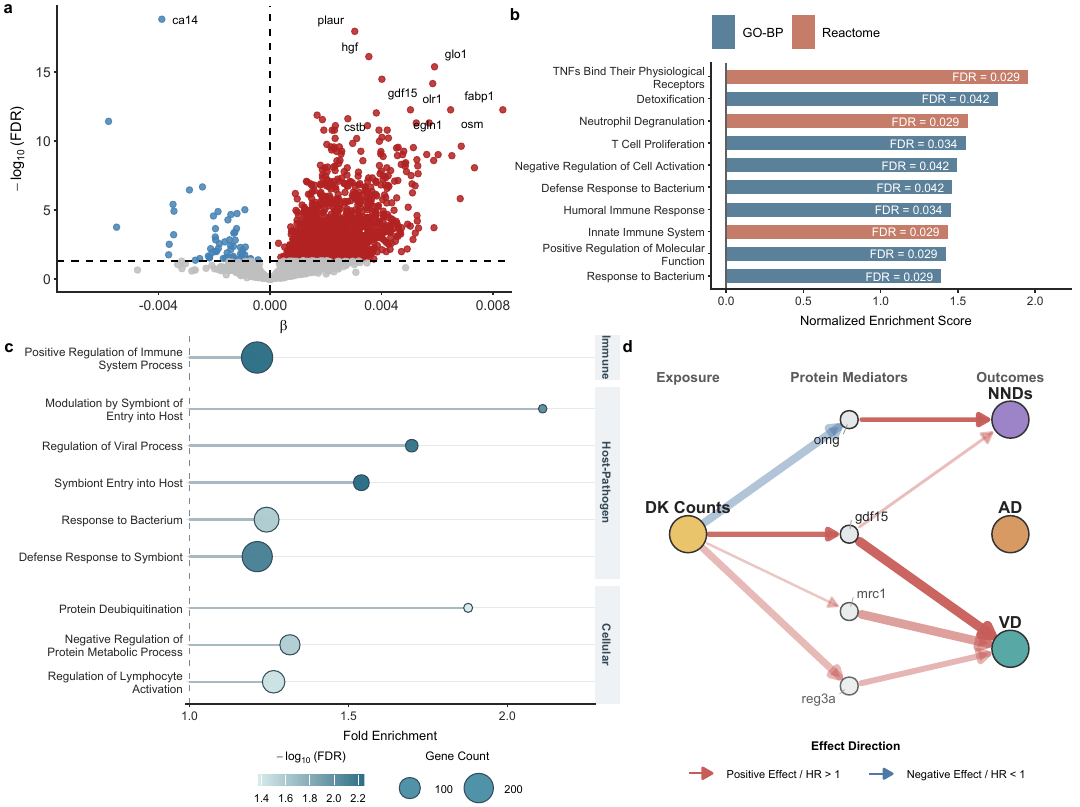}
    \caption{\textbf{Associations between DK response burden and incident
    chronic diseases.} The forest plot presents hazard ratios and 95\%
    confidence intervals across DK response groups for chronic ischaemic heart
    disease, hypertension, type 2 diabetes, disorders of lipoprotein
    metabolism, asthma, polyarthrosis, chronic obstructive pulmonary disease,
    chronic kidney disease, obesity, depressive disorders, and sleep
    disorders. Participants reporting 0--1 DK responses were the reference,
    and those with the corresponding diagnosis at or before baseline were
    excluded separately for each outcome. Cox models were adjusted for age,
    sex, BMI, educational attainment, Townsend Deprivation Index, ethnicity,
    household income, smoking, physical activity, sleep duration, and alcohol
    consumption. Squares show hazard-ratio estimates and horizontal lines show
    95\% confidence intervals.}
    \label{fig:5}
\end{figure*}

The questionnaire-domain and circulating-molecular analyses suggested that DK response burden may capture vulnerability extending beyond cognitive or neurodegenerative processes. We therefore examined its prospective associations with a broader range of incident chronic diseases. Participants with a diagnosis recorded on or before baseline were excluded separately for each outcome, and DK responses were categorised as 0--1, 2--4, 5--7, or \(>7\). Cox models were adjusted for demographic, socioeconomic, and behavioural characteristics (Fig.~\ref{fig:5}; \hyperref[method:chronic-outcomes]{Methods: Chronic-disease outcome analyses}). Complete estimates are reported in Supplementary Table S15.

A graded increase in risk across DK groups was observed for all 11 chronic-disease outcomes examined (Fig.~\ref{fig:5}). The largest associations were observed for incident depressive disorders and chronic obstructive pulmonary disease. Compared with participants reporting 0--1 DK responses, those reporting more than seven had a 57\% higher risk of depressive disorders (HR, 1.57; 95\% CI, 1.52--1.63) and a 46\% higher risk of chronic obstructive pulmonary disease (HR, 1.46; 95\% CI, 1.40--1.52). Associations were already evident in the intermediate DK groups, reaching HRs of 1.33 (95\% CI, 1.29--1.38) for depressive disorders and 1.25 (95\% CI, 1.20--1.31) for chronic obstructive pulmonary disease among participants with 5--7 DK responses.

Higher DK burden was also associated with incident cardiometabolic diseases. In the group with more than seven DK responses, the HR was 1.31 (95\% CI, 1.27--1.35) for type 2 diabetes, 1.19 (95\% CI, 1.17--1.22) for essential hypertension, 1.17 (95\% CI, 1.13--1.21) for chronic ischaemic heart disease, and 1.17 (95\% CI, 1.15--1.20) for disorders of lipoprotein metabolism. Corresponding associations were observed for chronic kidney disease (HR, 1.25; 95\% CI, 1.20--1.30), obesity (HR, 1.11; 95\% CI, 1.08--1.15), and sleep disorders (HR, 1.25; 95\% CI, 1.17--1.33).

The pattern extended to respiratory and musculoskeletal outcomes. Participants with more than seven DK responses had higher risks of asthma (HR, 1.25; 95\% CI, 1.21--1.29) and polyarthrosis (HR, 1.20; 95\% CI, 1.14--1.27). For nearly all outcomes, risk estimates increased progressively from the 2--4 group to the 5--7 and $>7$ groups, indicating that the chronic-disease associations were not restricted to the small proportion of participants with the most frequent DK responses.

The chronic-disease spectrum places the dementia findings within a wider phenotype spanning psychiatric, respiratory, metabolic, cardiovascular, renal, and musculoskeletal vulnerability. Together with the domain and molecular results, this breadth suggests that DK burden integrates cognitive difficulty with health literacy, socioeconomic disadvantage, multimorbidity, psychological health, and functional capacity.

\section*{Discussion}

In this large prospective cohort, responses conventionally treated as missing questionnaire data contained reproducible information about subsequent health risk. A higher burden of DK responses was socially and clinically patterned at baseline and was prospectively associated with higher risks of the dementia-related disease composite, AD, and VD. The associations were graded, persisted after extensive adjustment and removal of diagnoses within two years, and were reproduced directionally in the geographic replication sample. Domain decomposition localised part of the signal to uncertainty about household, medical, and personal information, while molecular and chronic-disease analyses placed it within a broader systemic phenotype. Questionnaire response uncertainty therefore emerges as a candidate early digital sign: a low-cost behavioural feature already embedded in electronic assessment that identifies dementia-related vulnerability before recorded diagnosis.

The first conceptual implication concerns how explicit uncertain responses are handled in epidemiological data. Conventional missing-data methods focus primarily on the statistical consequences of incomplete observations and on recovering valid estimates through approaches such as multiple imputation \cite{sterne2009multiple,pedersen2017missing,little2019statistical}. These methods remain essential when information is genuinely unavailable. However, an explicit DK selection is not equivalent to an unrecorded value caused by technical failure, loss to follow-up, or accidental omission. It is an observed response behaviour generated by the interaction between a participant, a question, and a digital assessment environment. Recoding all such responses directly as generic missing values may therefore discard information about recall, comprehension, confidence, health knowledge, engagement, or the complexity of the requested information. Our results do not imply that DK values should be substituted for the substantive answer to a question; rather, they suggest that the act of selecting DK can itself be retained as an analytically meaningful feature.

Several mechanisms may contribute to this signal. Difficulties retrieving autobiographical or health-related information could increase the probability of an uncertain response. Reduced confidence in memory, impaired monitoring of one's own knowledge, declining executive control, fatigue, low health literacy, and reduced engagement with lengthy assessments may have similar effects. This interpretation is conceptually related to research on subjective cognitive decline and early changes in cognitive self-appraisal \cite{jessen2014conceptual,rabin2017subjective,jessen2020characterisation}. Nevertheless, DK responses differ importantly from a direct complaint of cognitive decline: they arise during individual questionnaire items and may reflect cognitive and non-cognitive processes simultaneously. Socioeconomic resources, educational experience, language, cultural response norms, multimorbidity, psychological health, and familiarity with one's medical history can all influence whether a participant feels able to provide a definite answer. DK burden should therefore be viewed as a composite behavioural phenotype rather than a direct measure of memory impairment.

DK burden has three properties that make the early-digital-sign framing useful. It is generated passively during a routine assessment, available before a recorded clinical diagnosis, and derivable without additional devices, biological sampling, or dedicated cognitive testing. Unlike dedicated UK Biobank cognitive tests, for which psychometric validity and reliability have been evaluated, DK burden was not designed or validated as a cognitive test \cite{fawnsritchie2020}.  These properties accord with the broader role of digital health measurements in capturing subtle behavioural variation at low marginal cost \cite{allard2014mobile,coravos2019digital}. The prospective dose--response pattern, two-year lag analysis, and geographic replication connect this behavioural trace to later diagnosis, while the chronic-disease and molecular analyses show that it represents a wider form of behavioural and health uncertainty than disease-specific preclinical pathology \cite{jack2010hypothetical,sperling2011toward,dubois2016}.

The geographic split-sample analysis supports internal reproducibility beyond the larger England development sample. Participants recruited in Scotland and Wales showed the same direction and an elevated risk in the highest DK group for all three dementia-related outcomes. The attenuated lower-group estimates and evidence of regional heterogeneity for AD, together with the smaller replication event counts, indicate that effect magnitude may vary across settings. Independent cohorts with different questionnaires, languages, and health-record systems are needed to assess transportability beyond UK Biobank.

The PRS analyses further clarified what the DK signal represents. Polygenic
susceptibility to several vascular and cardiometabolic traits was associated
with DK burden, but even the strongest individual PRS explained less than
0.1\% of its variance. At the same time, disease-relevant PRSs showed the
expected associations with incident NNDs, AD, and VD, demonstrating that the
scores captured meaningful inherited susceptibility. Their inclusion changed
the DK log-HR by less than 3\%, and no multiplicative DK-by-PRS interaction
survived FDR correction. DK burden is therefore unlikely to be primarily a
proxy for the measured polygenic liabilities. Instead, questionnaire response
uncertainty and inherited susceptibility appear to contribute complementary
risk information. The pronounced risks among participants with both high DK
burden and high PRS illustrate the accumulation of these two dimensions,
without providing evidence of biological interaction on the multiplicative
scale.

The questionnaire-domain analysis sharpened the behavioural interpretation of the aggregate DK count. Household-related DK responses showed the strongest domain-specific associations with both AD and VD, while female-specific factors and medical-condition questions were shared across the two dementia outcomes. Medication, cancer-screening, and family-history domains contributed additional associations with AD, whereas electronic-device-use and chest-pain domains were associated with VD. These domains differ in content but share demands on the retrieval, monitoring, and organisation of everyday information. Household and family-history questions draw on autobiographical and contextual knowledge, medication and medical-condition questions require access to and integration of personal health information, and electronic-device-use questions additionally involve familiarity with routine functional activities. This pattern accords with longitudinal evidence that subtle mistakes, inconsistencies, skipped information, and other reductions in survey-response quality can precede dementia diagnosis and cognitive decline \cite{schneider2021subtle,schneider2022quality,schneider2024cognitive}. The recurrence of medical-condition and chest-pain DK responses across several chronic diseases further connects response uncertainty with multimorbidity and the cognitive and organisational demands involved in monitoring health. Taken together, the domain results suggest that cumulative DK burden becomes particularly informative when uncertainty occurs in questions that depend on recalling, integrating, and confidently reporting ordinary life and health information.

The molecular analyses provide a biological bridge between questionnaire response uncertainty and systemic processes previously implicated in dementia. The metabolomic signal was anchored by lower DHA and omega-3-related measures together with differences in HDL-related lipid traits. Prospective UK Biobank studies have independently associated circulating omega-3 and DHA measures, as well as fatty-acid, lipoprotein, ketone-body, glucose, and amino-acid profiles, with subsequent dementia, AD, or VD \cite{sala2023omega3,huang2023metabolites}. Routine biochemistry extended this convergence across nutritional, renal, glycaemic, and inflammatory systems: the lower vitamin D and creatinine-to-cystatin C ratio and the higher cystatin C, glycated haemoglobin, C-reactive protein, and inflammatory cell-count indices observed with increasing DK burden closely parallel laboratory profiles prospectively associated with incident dementia in UK Biobank \cite{he2022serum}. The proteomic results added a complementary immune and cellular-stress dimension. GDF15, one of the strongest DK-associated proteins, has independently emerged as a prominent prospective marker of all-cause dementia, AD, and VD \cite{guo2024proteomic}, while large-scale proteomic studies have linked future dementia to immune, lipid-metabolic, vascular, and haemostatic protein networks \cite{walker2021proteomic}. The convergence of these modalities indicates that higher DK burden accompanies a circulating state characterised by altered lipid metabolism, glycaemic and renal stress, nutritional differences, systemic inflammation, and peripheral immune activation---biological systems that repeatedly recur in prospective studies of dementia risk.

The exploratory mediation analyses further connected components of this molecular profile with subsequent disease outcomes. Valine was identified across NNDs, AD, and VD, while blood-marker indirect effects involved glycated haemoglobin, vitamin D, renal-function indices, C-reactive protein, and inflammatory cell counts; in the proteomic analysis, GDF15 contributed to both NND and VD models. Considered together with the association and pathway analyses, these results support a distributed molecular architecture in which metabolic, renal, inflammatory, and immune processes each account for a small component of the relationship between DK burden and later disease. Because DK responses and the circulating candidate mediators were measured at the same baseline assessment, the indirect-effect estimates describe statistical mediation rather than establishing temporal molecular causation, which will require repeated molecular measurements.

The chronic-disease findings place the dementia associations within a broader multimorbidity framework. Higher DK burden predicted depressive disorders, chronic obstructive pulmonary disease, type 2 diabetes, chronic kidney disease, sleep disorders, cardiovascular conditions, asthma, obesity, and polyarthrosis. Several of these conditions have themselves been prospectively linked to subsequent dementia. In the Whitehall II cohort followed for more than 30 years, multimorbidity was associated with substantially higher dementia incidence, with individual associations observed for diabetes, hypertension, chronic kidney disease, chronic obstructive pulmonary disease, depression, and arthritis, and with stronger risks as the number and severity of chronic conditions increased \cite{benhassen2022multimorbidity}. This complements established evidence that cardiovascular and metabolic disease, depression, obesity, physical inactivity, and related modifiable exposures contribute to dementia risk across the life course \cite{livingston2020dementia}. The chronic-disease spectrum observed here therefore connects DK burden to the same multimorbidity structure that precedes dementia in longitudinal studies. Frequent uncertain responses may integrate the cognitive and functional demands of living with multiple conditions together with differences in health knowledge, psychological health, socioeconomic resources, and the capacity to organise and report increasingly complex medical information.

These observations have implications for epidemiological analysis and digital health research. Explicit DK selections can be retained alongside the missing-data treatment required for the substantive variables, allowing response uncertainty to be summarised at item, domain, and questionnaire levels. Repeated questionnaires could reveal whether increasing uncertainty carries more information than a single baseline count. In population-health settings, externally validated DK features could also help prioritise more detailed assessment of cognition, psychological health, functional capacity, or unmet medical needs. Translation will require prespecified thresholds, calibration, decision-curve and fairness analyses, and evaluation across languages, cultures, and questionnaire designs.

This study has several strengths. The large UK Biobank sample provided substantial statistical power and enabled the evaluation of multiple incident outcomes using linked longitudinal health records. DK burden was defined before outcome ascertainment and examined using complementary categorical and continuous models. The analysis incorporated exposure--response modelling, a two-year lag analysis, subgroup assessment, geographic split-sample replication, and extensive adjustment for demographic, socioeconomic, behavioural, and clinical factors. Questionnaire-domain decomposition helped localise the behavioural signal, while NMR metabolomics, routine biochemistry, and plasma proteomics provided complementary information about its systemic correlates.

Several limitations must also be considered. First, this was an observational study, and residual confounding cannot be eliminated. DK burden may partly reflect unmeasured educational quality, literacy, language proficiency, sensory impairment, fatigue, personality, healthcare access, or baseline functional status. Reverse causation also remains possible because subtle cognitive or systemic disease processes may have been present before formal diagnosis, although the lag analysis reduced the likelihood that associations were driven entirely by immediately impending diagnoses. Second, DK burden was measured primarily at a single baseline assessment. Repeated questionnaires will be needed to determine whether increasing uncertainty over time is more informative than a single measurement.

Third, participants did not necessarily have equal opportunities to provide a DK response. Touchscreen branching logic, conditional questions, sex-specific items, and differences in question applicability can affect domain-specific counts. Although mutually adjusted domain models and FDR correction reduce some concerns, future analyses should explicitly incorporate the number of questions presented or eligible within each domain and should distinguish structural non-applicability from response uncertainty. Fourth, diagnoses were derived from hospital inpatient records and may miss milder, undiagnosed, or primary-care-managed cases.

Fifth, the UK Biobank is a large but selected volunteer cohort and is not fully representative of the general UK population \cite{sudlow2015uk}. The geographic split tested reproducibility within the same underlying study infrastructure rather than providing independent external validation. Generalisability to populations with different ancestry distributions, socioeconomic structures, languages, healthcare systems, or digital interfaces therefore remains uncertain. Changes in the apparent C-index were assessed in the same data used for model fitting and do not quantify out-of-sample prediction. Sixth, the molecular analyses were conducted in subsets with available biomarker measurements. Their very large sample sizes permitted small effects to reach statistical significance, and the observed associations may still reflect selection or residual confounding.

In conclusion, cumulative DK responses formed a graded, temporally robust, and geographically reproducible signal of subsequent dementia-related diagnoses. Questionnaire-domain, molecular, and chronic-disease analyses showed how this signal integrates cognitive, social, psychological, and systemic health vulnerability. Retaining response uncertainty as an observed behavioural feature converts information usually discarded as missing into a scalable early digital sign for dementia-risk research and population-level assessment.

\section*{Methods}
\phantomsection
\label{sec:methods}

\subsection*{Study design, population, and ethics}
\phantomsection
\label{method:study-population}

This prospective cohort study used data from the UK Biobank, a population-based resource that recruited more than 500,000 participants across England, Scotland, and Wales between 2006 and 2010 \cite{sudlow2015uk}. The present analysis included 502,234 participants with baseline touchscreen-questionnaire data and an available measure of DK responses.

At baseline, participants completed a touchscreen questionnaire, underwent physical measurements, and provided biological samples. Incident diagnoses were ascertained from linked hospital inpatient records; death-registry dates were used for censoring in the geographic and questionnaire-domain pipelines. Molecular analyses were performed in the corresponding UK Biobank subsets with available NMR metabolomic, blood-biochemistry, or plasma-proteomic measurements.

UK Biobank received ethical approval from the North West Multi-centre Research Ethics Committee (REC reference 16/NW/0274). All participants provided written informed consent. Data were accessed under UK Biobank application 146760 and were analysed in de-identified form.

\subsection*{Assessment of DK responses}
\phantomsection
\label{method:dk-assessment}

The primary exposure was the number of explicit DK responses recorded during the baseline touchscreen assessment. A catalogue of 396 baseline touchscreen fields was constructed from the UK Biobank field identifiers and questionnaire metadata. For each field, only the response code \texttt{-1}, corresponding to a ``Do not know'' response, was counted as DK. Other negative response codes, unavailable values, and \texttt{NA} values were not counted as DK. Thus, the exposure represented an observed response choice rather than generic item non-response.

The total DK count (\texttt{Counts\_DK}) was calculated by summing DK responses across the 396 baseline fields. For the primary analyses, participants were classified into four prespecified groups according to the empirical distribution of DK counts: 0--1, 2--4, 5--7, and $>7$ responses, with the 0--1 group used as the reference. DK count was also analysed as a continuous exposure to evaluate dose--response patterns.

The touchscreen fields were additionally classified into source questionnaire categories. The original catalogue contained 32 classes; the derived ``MET Scores'' class was excluded from domain-specific analyses because it did not represent a directly answered questionnaire domain, leaving 31 questionnaire domains. The field catalogue is provided in Supplementary Table S1, and total and domain-specific DK distributions are presented in Supplementary Fig.~S2 and Supplementary Table S2.

We distinguished explicit DK responses from conventional missing values because reasons for missingness and response generation may carry different information in epidemiological data \cite{sterne2009multiple,pedersen2017missing}. In this study, DK responses were retained as the exposure of interest and were not imputed.

\subsection*{Covariates and missing covariate data}
\phantomsection
\label{method:covariates}

Covariates were selected a priori on the basis of their potential relationships with questionnaire responding, socioeconomic circumstances, lifestyle, and subsequent disease risk. Sociodemographic covariates comprised age, sex, BMI, educational attainment, ethnicity, household income, and the Townsend deprivation index. Lifestyle covariates included smoking status, physical activity, sleep duration, and weekly alcohol consumption. Baseline health covariates comprised prevalent cardiovascular disease, depression, dyslipidaemia, diabetes, and hypertension.

Three nested adjustment models were used:

\begin{itemize}
    \item \textit{Model 1}: age, sex, BMI, educational attainment, Townsend deprivation index, ethnicity, and household income;
    \item \textit{Model 2}: Model~1 plus smoking status, physical activity, sleep duration, and weekly alcohol consumption;
    \item \textit{Model 3}: Model~2 plus prevalent cardiovascular disease, depression, dyslipidaemia, diabetes, and hypertension.
\end{itemize}

For the main association analyses, missing continuous covariates were replaced by the sample median, whereas missing categorical or binary covariates were replaced by the modal category. Molecular outcomes were not imputed; each molecular model was fitted among participants with an available value for that outcome. In the geographic replication analysis, imputation values and factor definitions were estimated exclusively in the England development sample and applied unchanged to the Scotland and Wales replication sample. Variable-level missingness is summarised in Supplementary Table S23.

\subsection*{Outcome ascertainment and follow-up}
\phantomsection
\label{method:outcomes}

Disease outcomes were identified from linked hospital inpatient records using the first recorded International Classification of Diseases, 10th Revision (ICD-10) diagnosis date. Participants with a diagnosis on or before their baseline assessment date were considered prevalent cases and were excluded from the corresponding incident-outcome analysis.

The primary outcome was a dementia-related disease composite denoted NNDs in the source analyses, figures, and tables; it comprised AD, VD, and other recorded dementias. AD was defined using ICD-10 codes F00.0, F00.1, F00.2, F00.9, G30.0, G30.1, G30.8, and G30.9. VD was defined using F01.0, F01.1, F01.2, F01.3, F01.8, and F01.9. Other dementia codes included F02.0--F02.4, F02.8, F02.9, and F03.0--F03.3 and F03.9. AD and VD were also analysed separately.

A motor-system disease composite was examined as an exploratory specificity comparison and comprised Parkinson's disease and amyotrophic lateral sclerosis. Parkinson's disease was identified using G20 and G21-series codes, and amyotrophic lateral sclerosis and related motor neurone diseases using G12-series codes. These outcomes were selected to compare associations for dementia-related and predominantly motor-system disorders.

The chronic disease analyses included coronary heart disease (I25), hypertension (I10), type 2 diabetes mellitus (E11), dyslipidaemia (E78), asthma (J45), polyarthrosis (M15), chronic obstructive pulmonary disease (J44), chronic kidney disease (N18), obesity (E66), depression (F32--F33), and sleep disorders (G47). Complete outcome definitions are provided in Supplementary Table S3.

For the primary Cox, restricted cubic spline, subgroup, two-year lag, and chronic-disease pipelines, participants without a recorded event were administratively censored on 1 January 2023. Follow-up time was calculated from the baseline assessment date and expressed in years using 365.25 days per year.

\subsection*{Baseline and descriptive analyses}
\phantomsection
\label{method:descriptive}

Baseline characteristics were summarised across the four DK groups. Continuous variables were reported as medians with interquartile ranges, and categorical variables as numbers and percentages. Group differences were evaluated using Kruskal--Wallis tests for continuous variables and chi-squared tests for categorical variables. Baseline characteristics are presented in Table~\ref{tab:baseline}.

The distribution of total DK counts was summarised using integer categories from 0 to 9 and a combined \(>9\) category. Questionnaire-level DK distributions were summarised across 32 source categories, including the derived ``MET Scores'' category, together with the aggregate count (Supplementary Table S2). Domain-specific Cox models excluded ``MET Scores'' and therefore used 31 directly answered questionnaire domains.

\subsection*{Primary prospective analyses}
\phantomsection
\label{method:primary}

Cox proportional hazards regression was used to estimate HRs and 95\% CIs for incident NNDs, AD, and VD across the four DK groups. The 0--1 group was the reference category. Models 1--3 were fitted sequentially to evaluate the influence of socioeconomic, lifestyle, and baseline health covariates. Tied event times were handled using the Efron approximation. Model~3 was the principal model, with complete estimates reported in Supplementary Table S4.

\phantomsection
\label{method:rcs}
To characterise the continuous dose--response relationship, restricted cubic spline models were fitted using the fully adjusted covariate set. DK counts above 10 were truncated at 10 to reduce the influence of the sparse upper tail. Four spline knots were used, and the sample median DK count was specified as the reference. Predicted HRs were generated between the 0.5th and 99.5th percentiles of the DK-count distribution. Overall and nonlinear associations were evaluated using the corresponding spline-model Wald statistics. The grouped and continuous analyses are presented in Fig.~\ref{fig:2}.
Spline coordinates and tests are reported in Supplementary Table S5.

\subsection*{Motor-system comparisons and two-year lag analyses}
\phantomsection
\label{method:negative-controls}

The grouped Cox models were repeated for the motor-system disease composite, Parkinson's disease, and amyotrophic lateral sclerosis using the same covariate hierarchy as the primary analysis. These results are presented in Supplementary Fig.~S1 and Supplementary Table S7.

To reduce the influence of diagnoses close to baseline, we performed a two-year lag analysis. Participants with a primary outcome diagnosis at baseline or during the first 730 days of follow-up were excluded. Follow-up time continued to be measured from the baseline assessment date, and Model~3 was refitted in the remaining population. Results are presented in Supplementary Fig.~S3 and Supplementary Table S6.

\subsection*{Geographic split-sample replication}
\phantomsection
\label{method:geographic-validation}

Geographic split-sample replication was conducted by separating assessment centres into an England development sample and a prespecified Scotland and Wales replication sample. Scottish assessment centres were identified by centre codes 11004 and 11005, and Welsh centres by codes 11003, 11022, and 11023. Participants recruited at all remaining English centres formed the development sample.

All preprocessing parameters, including continuous-variable medians, categorical modes, and factor levels, were estimated in the England sample and frozen before application to the Scotland and Wales sample. The latter sample was not used to derive preprocessing rules or the analysis specification; region-specific Cox models were then fitted separately within both samples.

Models 1--3 were fitted separately in the development and geographic replication samples. Heterogeneity between regions was examined using joint tests of the interaction between DK group and geographic sample. A centre-stratified Model~3 was examined as sensitivity analysis. Proportional hazards assumptions were evaluated using tests based on scaled Schoenfeld residuals \cite{schoenfeld1982}.

Geographic cohort construction, complete association estimates, interaction tests, and sensitivity analyses are reported in Table~\ref{tab:geographic-validation} and Supplementary Tables S8--S11.

\subsection*{Subgroup and interaction analyses}
\phantomsection
\label{method:subgroups}

Potential effect modification of the association between DK group and incident NNDs was explored across strata defined by age ($<50$, 50--60, and $>60$ years), sex, BMI ($\leq24.9$ and $>24.9$ kg/m$^{2}$), Townsend deprivation tertile, ethnicity, household income, educational attainment, smoking status, physical activity, sleep-duration category, alcohol consumption, and baseline cardiovascular disease, depression, dyslipidaemia, diabetes, and hypertension.

Within each subgroup, Model~3 was fitted after excluding a covariate when it defined the subgroup itself. Formal interaction was evaluated by comparing models with and without the DK-group-by-subgroup interaction using likelihood-ratio tests. Subgroup estimates and interaction tests are reported in Supplementary Tables S12 and S13.

\subsection*{Polygenic risk-score analyses}
\phantomsection
\label{method:prs}

We evaluated 39 Standard and 51 Enhanced UK Biobank PRSs. The Enhanced scores
were available in a smaller participant subset and were therefore treated as
complementary sensitivity analyses. Total DK count and each PRS were
standardised to a mean of zero and a standard deviation of one within the
corresponding analysis sample.

Associations between individual PRSs and standardised DK burden were estimated
using linear regression adjusted for age, sex, BMI, educational attainment,
Townsend deprivation index, self-reported ethnicity, and household income. The
incremental contribution of each PRS was summarised using the change in
\(R^{2}\), partial \(R^{2}\), and a likelihood-ratio test comparing models
with and without the PRS. Benjamini--Hochberg FDR correction was applied
separately within the Standard and Enhanced PRS sets. Complete results are
reported in Supplementary Table S24.

For the prospective analyses, we evaluated eight biologically motivated outcome–PRS pairs: AD and cardiovascular disease PRSs for the NNDs composite; the AD PRS for AD; and cardiovascular disease, ischaemic stroke, hypertension, atrial fibrillation, and AD PRSs for VD. The AD PRS–VD pair was included to examine potentially shared dementia susceptibility. Each pair
was evaluated using both the Standard and Enhanced PRS where available. For
each disease--PRS pair, three fully adjusted Model~3 Cox models were compared:
a model containing standardised DK count, a model additionally containing the
standardised PRS, and a model additionally containing the DK-by-PRS
interaction. The percentage change in the DK log-HR after PRS adjustment,
likelihood-ratio tests, and the change in the apparent Harrell C-index \cite{harrell1996} were
calculated.

For joint-risk analyses, high DK burden was defined as more than seven DK
responses, and high and low polygenic risk were defined as the highest and
lowest PRS quintiles, respectively. Joint categories were entered into Model~3
Cox regression, with participants reporting no more than seven DK responses
and belonging to the lowest PRS quintile as the reference group. Proportional
hazards assumptions were evaluated using scaled Schoenfeld residuals.
Benjamini--Hochberg FDR correction was applied to the targeted PRS,
interaction, and joint-risk tests. Complete estimates and diagnostic results
are reported in Supplementary Table S25.

\subsection*{Questionnaire-domain-specific analyses}
\phantomsection
\label{method:questionnaire-domains}

To identify which components of the touchscreen assessment contributed most strongly to the observed associations, DK counts were calculated separately for each of the 31 directly answered questionnaire domains. Domains with no variation were excluded. All estimable domain-specific DK counts were entered simultaneously into the same fully adjusted Cox model so that each coefficient represented the association per additional DK response within that domain, conditional on DK responses in the other domains.

Domain-specific associations were estimated for AD, VD, coronary heart disease, hypertension, type 2 diabetes, dyslipidaemia, asthma, and polyarthrosis. The Benjamini--Hochberg procedure was applied across all estimable domain--outcome associations. Results are presented in Fig.~\ref{fig:6}, with complete estimates in Supplementary Table S14.

\subsection*{Chronic disease analyses}
\phantomsection
\label{method:chronic-outcomes}

Associations between DK groups and incident chronic diseases were evaluated using Cox regression. Participants with the corresponding condition at or before baseline were excluded. These analyses used Model~2, incorporating sociodemographic and lifestyle covariates without additionally adjusting for the other baseline chronic diseases, because several could lie on shared or downstream disease pathways. HRs and 95\% CIs for the 2--4, 5--7, and $>7$ groups were estimated relative to the 0--1 group. Results are presented in Fig.~\ref{fig:5} and Supplementary Table S15.

\subsection*{NMR metabolomic analyses}
\phantomsection
\label{method:nmr-metabolomics}

Associations between continuous DK count and 251 plasma NMR metabolomic measures were examined among 212,853 participants with available NMR data. Each metabolite was analysed separately. The metabolite was standardised to a mean of zero and standard deviation of one within its analysis-specific sample and regressed on continuous DK count with adjustment for the full Model~3 covariate set.

Participants with a missing value for the relevant metabolite were excluded from that model. Covariates without variation in an analysis-specific sample were omitted. The regression coefficient therefore represented the standard-deviation difference in the metabolite per one additional DK response. Benjamini--Hochberg correction was applied across the 251 metabolite associations. Results are presented in Fig.~\ref{fig:3} and Supplementary Table S16.

\subsection*{Blood-biochemistry analyses}
\phantomsection
\label{method:blood-biochemistry}

Continuous DK count was examined in relation to 37 circulating biochemical and haematological measures covering lipid, liver, kidney, metabolic, hormonal, and inflammatory systems. In addition to directly measured biomarkers, the following derived indices were calculated: neutrophil-to-lymphocyte ratio, platelet-to-lymphocyte ratio, lymphocyte-to-monocyte ratio, systemic immune-inflammation index, systemic inflammation response index, creatinine-to-cystatin-C ratio, and triglyceride--glucose index. Values with invalid denominators or non-positive logarithmic arguments were treated as unavailable.

Each blood measure was standardised within the corresponding analysis sample and analysed using linear regression adjusted for Model~3 covariates. Models required at least 50 participants with an available outcome. Coefficients represented the standard-deviation difference in the blood measure per one additional DK response. Benjamini--Hochberg correction was applied across the 37 measures. Results are shown in Fig.~\ref{fig:3} and Supplementary Table S18.

\subsection*{Proteomic association analyses}
\phantomsection
\label{method:proteomic-associations}

Plasma proteomic analyses included 2,923 Olink protein measures from the UK Biobank proteomics resource \cite{sun2023proteomics}. Protein abundance was modelled as a function of continuous DK count and the full Model~3 covariate set. Protein-wise linear models were fitted using the \texttt{limma} framework \cite{ritchie2015limma}, followed by empirical-Bayes moderation of the standard errors. The coefficient represented the difference in normalised protein abundance per one additional DK response. Benjamini--Hochberg correction was applied across all 2,923 proteins, and proteins with FDR-adjusted \(P<0.05\) were considered DK-associated. Protein-level results are presented in Fig.~\ref{fig:4} and Supplementary Table S20.

\subsection*{Proteomic pathway analyses}
\phantomsection
\label{method:pathway-analysis}

Proteins associated with DK count at FDR-adjusted \(P<0.05\) were mapped from protein symbols to Entrez gene identifiers. Enrichment analyses used \texttt{clusterProfiler} \cite{yu2012clusterprofiler}. Over-representation analyses were performed for Gene Ontology biological process, molecular function, and cellular component categories, together with Kyoto Encyclopedia of Genes and Genomes (KEGG) and Reactome pathways. All successfully mapped proteins tested in the proteomic analysis were used as the background universe.

In parallel, gene-set enrichment analysis was conducted using all tested proteins ranked by the moderated \(t\) statistic. Gene sets were obtained from the Molecular Signatures Database (MSigDB), including Gene Ontology biological process, KEGG MEDICUS, and Reactome collections \cite{liberzon2011msigdb,msigdb2023keggmedicus}. The ranked-set analysis followed the gene-set enrichment framework \cite{subramanian2005gsea}. Pathways containing 15--500 mapped genes were tested using 10,000 permutations. Results with FDR-adjusted \(P<0.05\) were considered statistically significant. Pathway results are presented in Fig.~\ref{fig:4} and Supplementary Table S21.

\subsection*{Exploratory molecular mediation analyses}
\phantomsection
\label{method:indirect-effects}

Exploratory regression-based mediation analyses estimated model-based direct and indirect effects for DK count and incident NNDs, AD, or VD, with selected molecular measures as candidate mediators \cite{vanderweele2011survival,valeri2013mediation}.

Candidate NMR metabolites and blood biomarkers were selected from measures associated with DK count after FDR correction. DK count was contrasted between its 25th and 75th percentiles. A linear mediator model and a Cox outcome model were fitted with the full Model~3 covariate set. No exposure--mediator interaction was specified. Pure natural direct effects, total natural indirect effects, and total effects were estimated with 95\% CIs. Analyses required at least 500 participants and 30 outcome events. Benjamini--Hochberg correction was applied to indirect-effect tests within each disease outcome. Results are presented in Fig.~\ref{fig:3} and Supplementary Tables S17 and S19.

\subsubsection*{Protein mediation analysis}
\phantomsection
\label{method:protein-indirect-effects}

For the proteomic mediation analysis, the 30 proteins with the smallest FDR-adjusted \(P\) values in the DK--protein association analysis were carried forward. The same 25th-to-75th-percentile DK contrast, covariate set, linear mediator model, Cox outcome model, minimum sample-size criteria, and outcome-specific FDR correction were used. Protein indirect-effect estimates are presented in Fig.~\ref{fig:4} and Supplementary Table S22.

\subsection*{Statistical analysis and software}
\phantomsection
\label{method:statistical-analysis}

All statistical tests were two-sided. HRs, regression coefficients, and indirect-effect estimates are reported with 95\% CIs. A nominal \(P<0.05\) was used for primary and interaction analyses unless otherwise stated. For high-dimensional questionnaire-domain, metabolomic, blood-biochemistry, proteomic, and mediation analyses, the Benjamini--Hochberg procedure \cite{benjamini1995} controlled the FDR. Statistical significance was defined as FDR-adjusted \(P<0.05\).

No formal sample-size calculation was performed because all eligible UK Biobank participants with the required exposure and outcome data were included. Analyses were conducted with R version 4.5.1 in the RStudio environment. Principal packages included \texttt{dplyr}, \texttt{survival}, \texttt{rms}, \texttt{tableone}, \texttt{broom}, \texttt{limma}, \texttt{clusterProfiler}, \texttt{ReactomePA}, \texttt{msigdbr}, \texttt{fgsea}, \texttt{regmedint}, \texttt{ggplot2}, and \texttt{patchwork}.

\clearpage
\section*{Data availability}

The participant-level data used in this study are held by UK Biobank and cannot be redistributed by the authors. Eligible researchers may apply to UK Biobank for access through its research-access procedure (\url{https://www.ukbiobank.ac.uk/use-our-data/apply-for-access/}); this study used data under application 146760. Aggregate results and model estimates supporting the manuscript are provided in the accompanying Supplementary Tables workbook.

\section*{Code availability}
The R scripts used for data processing, statistical analyses, and figure generation are publicly available at \href{https://github.com/PJSSOD/UKB-Missing-Data}{https://github.com/PJSSOD/UKB-Missing-Data}. The repository contains no participant-level UK Biobank data. Access to the data required to reproduce the analyses is subject to approval by UK Biobank.

\section*{Acknowledgements}
We thank the Shanghai Institute for Mathematics and Interdisciplinary Sciences (SIMIS) for its financial support. This research was funded by SIMIS under grant number SIMIS-ID-2025-OM. The authors are grateful for the resources and facilities provided by SIMIS, which were essential to the completion of this work. Zhigang Yao was supported by the Singapore Ministry of Education through Tier 2 grants A-0008520-00-00 and A-8001562-00-00 and Tier 1 grant A-8004146-00-00 at the National University of Singapore. Yuqing Xia was supported by the National Natural Science Foundation of China (grant No. 12401500).

\section*{Author contributions}
Y.L. and B.L. contributed equally to this work. Y.L. contributed to literature collection, selection of the statistical analyses used in the study, implementation and debugging of the analysis code, data curation, preparation of the figures and tables, and drafting of the manuscript. B.L. contributed to literature collection, project planning and study design, selection of the statistical analyses used in the study, validation and interpretation of the results, design of the figures and tables, and structuring and writing of the manuscript. Y.X. and Z.Y. developed the study idea, supervised the project, obtained funding, contributed to interpretation of the results, and reviewed and revised the manuscript. All authors read and approved the final manuscript.

\section*{Ethics declaration}
UK Biobank has Research Tissue Bank approval from the North West Multi-centre Research Ethics Committee (REC reference 21/NW/0157). This approval covers the present secondary analysis, which was conducted under UK Biobank application 146760. All participants provided written informed consent.

\section*{Competing interests}
The authors declare no competing interests.


\end{document}


\maketitle

\section*{Supplementary Texts}

\subsection*{Overview of the UK Biobank Touchscreen Questionnaire}
\label{appendix:touchscreen}

At the baseline Assessment Centre visit, UK Biobank participants completed a
self-administered touchscreen questionnaire that collected standardised
information without interviewer-assisted data entry. The questionnaire employed
conditional branching, such that the sequence of questions presented to each
participant depended on their preceding responses. This design reduced exposure
to irrelevant questions but also resulted in participant-specific opportunities
to encounter individual items. The present study included 396 baseline
touchscreen fields that permitted an explicit ``Do not know'' (DK) response.
Field identifiers, question descriptions, and questionnaire-domain assignments
are provided in Supplementary Table~S1 \cite{ukb_touchscreen}.

For descriptive summaries, these fields were organised into 32 source
categories according to their subject matter and response context. The exact
grouping used in Supplementary Tables S1 and S2 was:

\begin{itemize}
    \item \textbf{Early life}: Early life factors.
    \item \textbf{Family history}: Family history.
    \item \textbf{Health and medical history}: Breathing; Cancer screening; Chest pain; Claudication and peripheral artery disease; Eyesight; General health; Hearing; Medical conditions; Medication; Mouth; Operations; and Pain.
    \item \textbf{Lifestyle and environment}: Alcohol; Diet; Electronic device use; MET Scores; Physical activity; Sexual factors; Sleep; Smoking; and Sun exposure.
    \item \textbf{Psychosocial factors}: Mental health and Social support.
    \item \textbf{Sex-specific factors}: Female-specific factors and Male-specific factors.
    \item \textbf{Sociodemographic factors}: Education; Employment; Ethnicity; Household; and Other sociodemographic factors.
\end{itemize}

``MET Scores'' was a derived category rather than a directly answered domain;
it was retained in descriptive displays but excluded from domain-specific Cox
models, leaving 31 analytic domains.

Collectively, these categories cover personal history, social circumstances,
lifestyle, environmental exposures, symptoms, and health knowledge. The
aggregate DK count therefore captures response uncertainty arising across
multiple components of the questionnaire response process. The domain-specific
analyses assessed whether the prospective associations with dementia-related
outcomes were broadly distributed across the questionnaire or concentrated
within particular types of information. Supplementary Table~S2 provides the
corresponding numerical distributions, whereas Supplementary
Fig.~\ref{fig:S2} visualises DK counts overall and within each questionnaire
source category.

\section*{Supplementary Figures}

\renewcommand{\thefigure}{S\arabic{figure}}
\setcounter{figure}{0}

\begin{figure}[H]
    \centering
    \includegraphics[width=\linewidth]{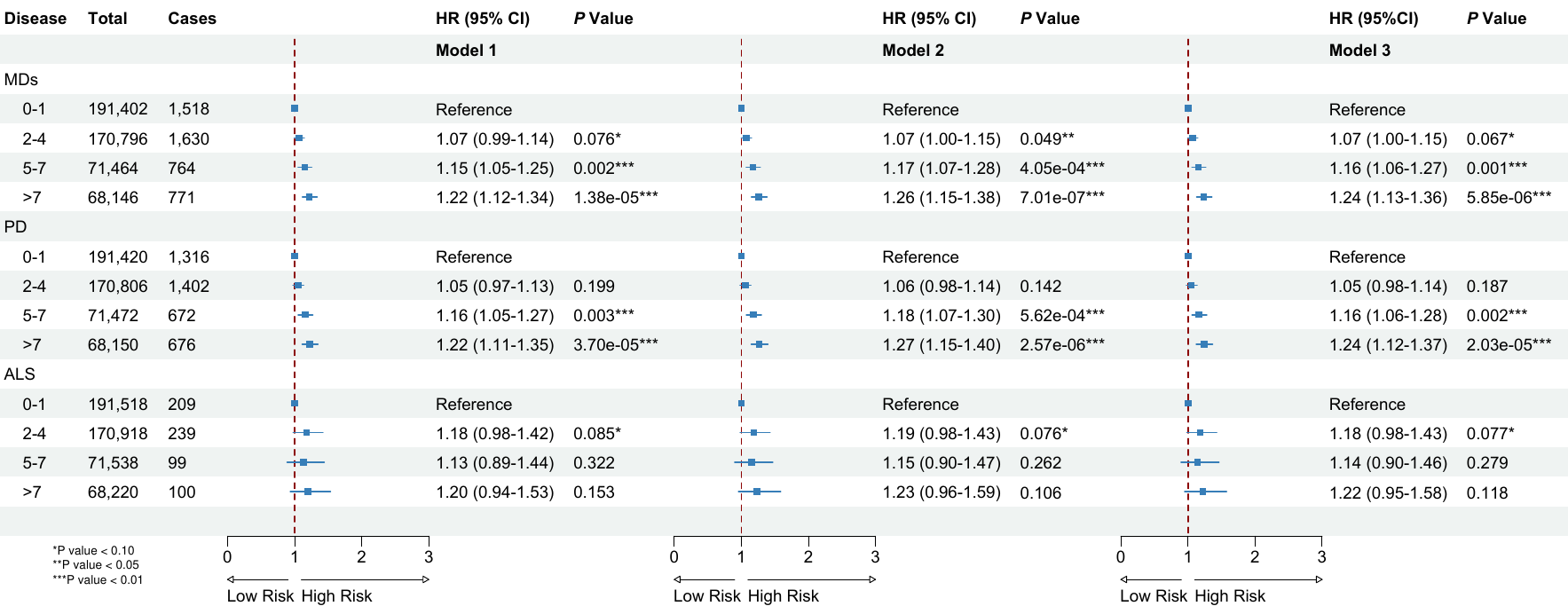}
    \caption{\textbf{Associations of DK response burden with incident
    motor-system degenerative outcomes.}
    Hazard ratios and 95\% confidence intervals are presented for the
    motor-system degenerative disease composite, Parkinson's disease and
    secondary parkinsonism, and amyotrophic lateral sclerosis and related
    motor-neurone disorders across categories of cumulative DK responses.
    Estimates are shown for the three progressively adjusted Cox models used
    in the primary analyses. Participants with 0--1 DK responses constituted
    the reference group.}
    \label{fig:S1}
\end{figure}

\clearpage
\begin{figure}[H]
    \centering
    \includegraphics[width=\textwidth]{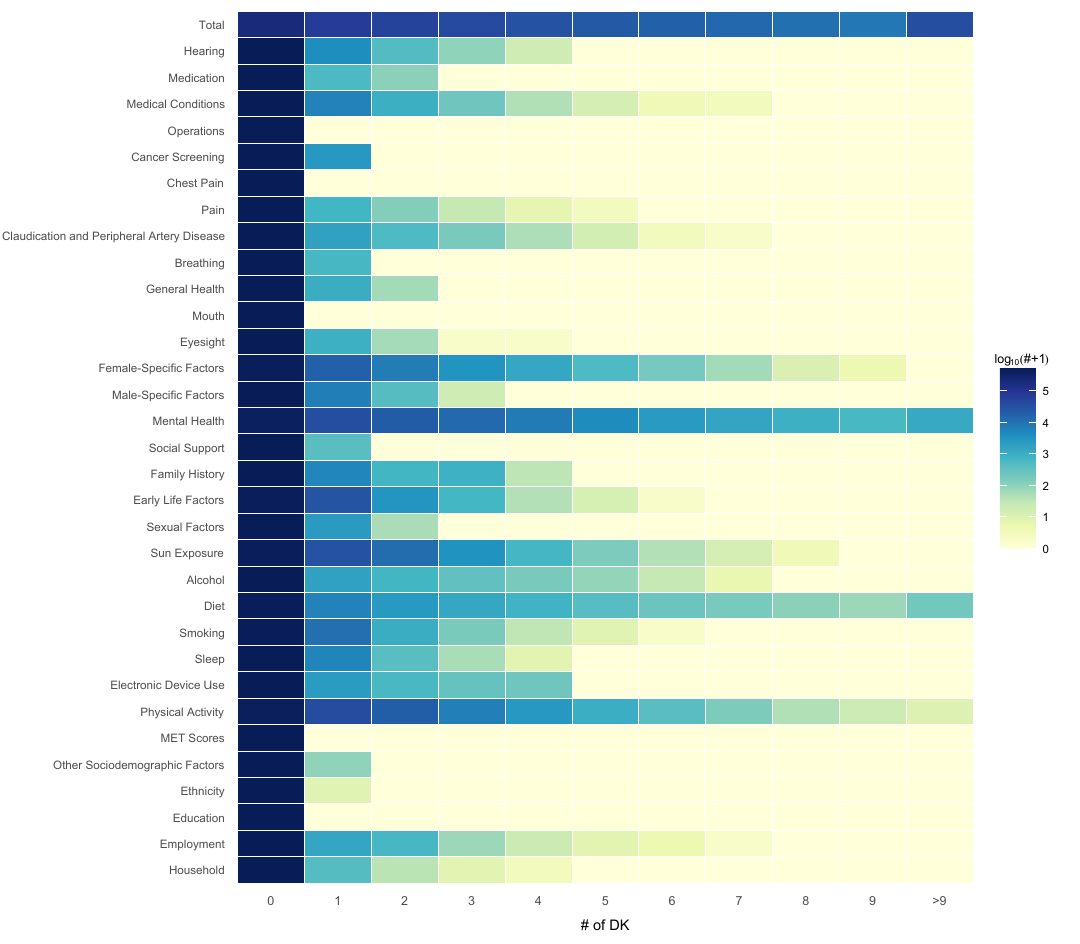}
    \caption{\textbf{Distribution of DK responses across UK Biobank
    touchscreen questionnaire domains.}
    Rows represent the 32 source categories and the aggregate count across
    all 396 eligible touchscreen fields. Columns indicate the number of DK
    responses recorded within each source category or across the complete touchscreen
    assessment. Each cell represents the number of participants with the
    corresponding DK count, with colour intensity scaled as
    \(\log_{10}(n+1)\), where \(n\) denotes the participant count.}
    \label{fig:S2}
\end{figure}

\begin{figure}[H]
    \centering
    \includegraphics[width=0.62\linewidth]{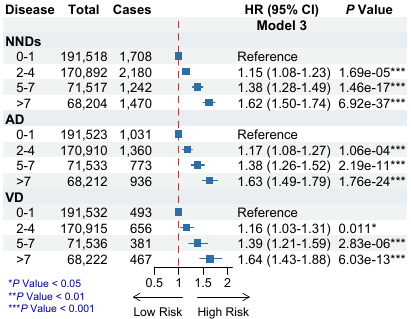}
    \caption{\textbf{Two-year lag sensitivity analysis of DK response burden
    and incident dementia-related outcomes.}
    Fully adjusted hazard ratios and 95\% confidence intervals are shown for
    the dementia-related disease composite, Alzheimer's disease, and vascular
    dementia after excluding diagnoses recorded at baseline or during the
    first 730 days of follow-up. Follow-up time remained anchored at the
    baseline assessment, and covariate adjustment was identical to Model~3 of
    the primary analysis. Participants with 0--1 DK responses constituted the
    reference group.}
    \label{fig:S3}
\end{figure}

\renewcommand{\refname}{Supplementary References}